\documentclass[twocolumn,english,aps,prl,reprint,nofootinbib,superscriptaddress]{revtex4-2}
\usepackage[T1]{fontenc}
\usepackage[latin9]{inputenc}
\usepackage{amsmath}
\usepackage{amssymb}
\usepackage{graphicx}
\usepackage{rotating}
\usepackage{tabularx}
\usepackage{braket}
\usepackage{color}
\usepackage{float}
\usepackage{hyperref}
\usepackage{multirow}
\hypersetup{
    colorlinks,
    linkcolor={black},
    citecolor={black},
    urlcolor={blue}
}

\makeatletter
\@ifundefined{textcolor}{}
{%
 \definecolor{BLACK}{gray}{0}
 \definecolor{WHITE}{gray}{1}
 \definecolor{RED}{rgb}{1,0,0}
 \definecolor{GREEN}{rgb}{0,1,0}
 \definecolor{BLUE}{rgb}{0,0,1}
 \definecolor{CYAN}{cmyk}{1,0,0,0}
 \definecolor{MAGENTA}{cmyk}{0,1,0,0}
 \definecolor{YELLOW}{cmyk}{0,0,1,0}
 \definecolor{PURPLE}{rgb}{0.7,0,0.7}
 \definecolor{dgreen}{rgb}{0,0.6,0}
}
\usepackage{xspace}
\newcommand{\A}{$A^2\Pi_{1/2}$\xspace}
\newcommand{\As}{$A^2\Pi$\xspace}
\newcommand{\B}{$B^2\Sigma^+$\xspace}
\newcommand{\X}{$X^2\Sigma^+$\xspace}
\newcommand{\cm}{cm$^{-1}$\xspace}
\newcommand{\Ppd}{$P_{\mathrm{pd}}$\xspace}

\usepackage{pslatex}

\usepackage{babel}

\makeatother

\begin{document}
\renewcommand{\arraystretch}{1.5}
\title{Probing the limits of optical cycling in a predissociative diatomic molecule}

\author{Qi Sun}
\affiliation{Department of Physics, Columbia University, New York, NY 10027-5255, USA}

\author{Claire E. Dickerson}
\affiliation{Department of Chemistry and Biochemistry, University of California, Los Angeles, California 90095, USA}

\author{Jinyu Dai}
\affiliation{Department of Physics, Columbia University, New York, NY 10027-5255, USA}

\author{Isaac M. Pope}
\affiliation{Department of Physics, Columbia University, New York, NY 10027-5255, USA}

\author{Lan Cheng}
\affiliation{Department of Chemistry, The Johns Hopkins University, Baltimore, Maryland 21218, USA}

\author{Daniel Neuhauser}
\affiliation{Department of Chemistry and Biochemistry, University of California, Los Angeles, California 90095, USA}

\author{Anastassia N. Alexandrova}
\affiliation{Department of Chemistry and Biochemistry, University of California, Los Angeles, California 90095, USA}

\author{Debayan Mitra}
\email{dm3710@columbia.edu}
\affiliation{Department of Physics, Columbia University, New York, NY 10027-5255, USA}

\author{Tanya Zelevinsky}
\affiliation{Department of Physics, Columbia University, New York, NY 10027-5255, USA}

\date{\today}

\begin{abstract}
\noindent 
Molecular predissociation, the spontaneous nonradiative bond breaking process, can limit the ability to scatter a large number of photons required to reach the ultracold regime in laser cooling. Unlike rovibrational branching, predissociation is irreversible since the fragments fly apart with high kinetic energy. {Of particular interest is the simple diatomic molecule, CaH, for which the two lowest electronically excited states used in laser cooling, \A and \B, lie above the dissociation threshold of the ground potential.} In this work, we present measurements and calculations that quantify the predissociation probabilities \Ppd affecting the cooling cycle. {For the lowest vibrational levels, we find \Ppd of $\sim10^{-6}$ for $A(v'=0)$ and $\sim10^{-3}$ for $B(v'=0)$.}  The results allow us to design a laser cooling scheme that will enable the creation of an ultracold and optically trapped cloud of CaH molecules. In addition, we use the results to propose a two-photon pathway to controlled dissociation of the molecules in order to gain access to their ultracold fragments, including hydrogen.
\end{abstract}
\maketitle

\section{Introduction}
\label{sec:Intro}

Rapid and repeated photon scattering is not only an efficient method of removing entropy from an atom or a molecule via photon recoils \cite{Metcalf_1999_laser_cooling_book}, but it also enables the high-fidelity single quantum state preparation and measurement needed for quantum information protocols \cite{Myerson_ion_qubit_spam,Blatt2012}. Optical cycling between the ground state and a low-lying electronic excited state, {pioneered with SrF \cite{Shuman_LaserCoolingDiatomic_2010} and CaF \cite{Truppe_DopplerLimitMolCooled_2017,Anderegg_2018_grey_molasses_CaF}}, has led to recent progress with laser cooled molecules such as tweezer arrays of CaF \cite{Anderegg_2019_CaF_tweezers}, a three-dimensional lattice of YO \cite{Wu_2021_YOlattice}, magneto-optical trapping (MOT) of CaOH \cite{Vilas_2022_3D_MOT_CaOH}, and one-dimensional Sisyphus cooling of CaOCH$_3$ \cite{Mitra_CaOCH3Sisphus_2020}. 

The primary challenge of direct laser cooling is the {large}
photon budget necessary for bringing a cryogenically precooled molecular beam to within the MOT capture velocity \cite{Hemmerling_2016,Williams_2017}. For example, typical molecular beams emanating from a cryogenic buffer gas beam (CBGB) source travel at mean forward velocities of $\sim200$~m$/$s \cite{Hutzler_CR2012_BufferGasBeams}. The recoil velocity per photon is $\sim2$~cm$/$s{, hence $>10^4$} photon scatters are needed to bring the molecular beam to a standstill. {The} photons {must} be scattered
faster than $10^6$~s$^{-1}$ to accomplish slowing within a $\sim1$~m distance. Satisfying
these criteria
can be challenging for molecules with complex internal structures. Indeed, alternative slowing schemes such as traveling wave Stark deceleration \cite{Aggarwal_2021_SrF_Stark_decelerator}, the electro-optic Sisyphus effect \cite{Zeppenfeld_2012_Sisyphus}, centrifuge deceleration \cite{Wu_2017_cryofuge}, and Zeeman-Sisyphus slowing \cite{Augenbraun_2021_Zeeman_Sisyphus} have been demonstrated.
These alternative schemes leverage state-dependent electric and magnetic field dependencies to remove entropy with minimal photon scatters. However, quantum-state resolved detection still requires optical cycling. 

\begin{figure*}[ht!]
   \centering
    \includegraphics[scale=1]{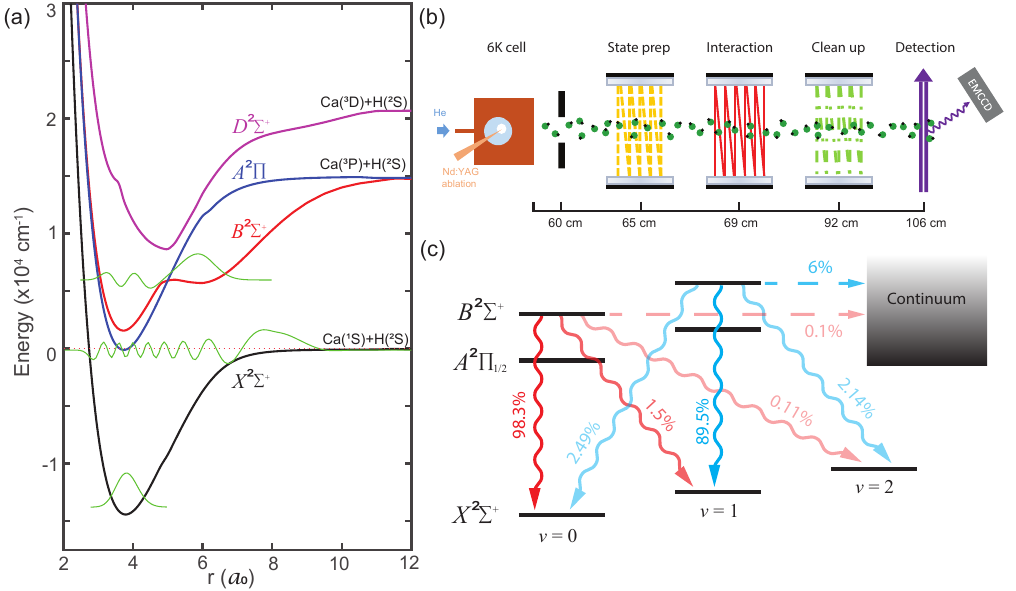}
    \caption{CaH molecular properties relevant to this work. (a) Potential energy surfaces (PES) for the 4 lowest electronic states:  \X, $A^2\Pi$, \B, and $D^2\Sigma^+$. Spin-orbit interaction is {omitted}.
    The $x$-axis is the internuclear separation $r$ in Bohr radii ($a_0$) and the $y$-axis is energy in cm$^{-1}$ (1~cm$^{-1}\approx$ 30~GHz). The energy origin is chosen as the Ca$(^1S)$+H$(^2S)$ continuum threshold ($v_{\mathrm{th}}$). Superimposed are the wavefunctions (bottom to top) for the $X(v''=0)$ absolute ground state, $X(v''=15)$
    {least-}bound state, and $B(v'=4)$ excited state. (b) Experimental layout used
    {in} this work. A buffer-gas cooled molecular beam emanates from the cryogenic cell and encounters 4 spatially separated regions: state preparation (S), interaction (I), cleanup (C), and detection (D).
    Each region {includes}
    multipassed lasers described in the text. {The diagram is not aligned to scale.} (c) Relevant vibrational branching ratios (squiggly arrows) calculated for the \B state. {The hyperfine structure of the excited states is unresolved}. Measured predissociation probabilities for $B(v'=0)$ and $B(v'=1)$ are denoted by dashed lines.}
    \label{fig:CaH_PES}
\end{figure*}
Although calcium monohydride (CaH) was among the earliest candidates proposed for laser cooling \cite{DiRosaEPJD04_LaserCoolingMolecules},
experimental progress was made {only}
recently \cite{Vasquez-Carson_2022_CaH}. One of the reasons is the unique electronic structure of CaH {compared to alkaline-earth monohalides} \cite{Gao_Summary_2014}. In CaH, the lowest-energy excited state {\A$(v'=0)$} lies
556~\cm above the Ca$(^1S)$+H$(^2S)$ dissociation threshold of the ground \X manifold (Fig.~\ref{fig:CaH_PES}(a)), so a molecule in the excited state {could}
decay into the continuum via a radiationless transition.
This phenomenon, known as predissociation \cite{Herzberg_molecular_spectra,Demtroder_molecular_physics}, {is traditionally studied by observing spectral line shapes and widths inconsistent with radiative decay}.
{A} predissociated molecule cannot be
repumped
{into optical cycling} because of the significant physical separation and relative velocity of the
fragments. Hence the predissociation probability (\Ppd) sets a limit on the number of photons one can scatter with laser cooling. 

Despite the fact that the \As state in CaH lies above the ground state threshold energy, predissociation from \As to the \X continuum is nominally forbidden due {to} the von Neumann-Wigner {non}crossing rule \cite{wigner_von_neumann}.
{For} diatomic molecules, states with different symmetries cross while those with the same symmetries form avoided crossings \cite{Teller1937,Mead1979}
{(i.e.,} the molecular Hamiltonian does not couple states with different symmetries).
{The} second-lowest excited \B state is {allowed}
to predissociate.
However, effects {such as}
spin-orbit coupling can lead to mixing of \As and \B states {resulting in}
a small but finite \Ppd for \As.  {Both $A$ and $B$ states are important for efficient optical cycling.}

In this work, we present theoretical analysis and measurements of predissociation probability for the \B state of CaH. We perform \textit{ab initio} calculations of the potential energy surfaces {for}
CaH,
and confirm their accuracy by extracting the Franck-Condon factors (FCFs) for the primary {\A$\rightarrow$ \X and \B$\rightarrow$ \X} decays and comparing them to our previous measurements. We calculate a nonradiative decay rate, and
{obtain an estimate of \Ppd}
by comparing it to the radiative decay rate. Next, we present a novel experimental protocol to measure an upper bound of \Ppd. We find that \Ppd$\approx 1\times10^{-3}$ for the vibrational ground state $(v'=0)$ and $\approx 6\times10^{-2}$ for the first vibrationally excited state $(v'=1)$ of the \B manifold. We deduce that the vibrational ground state of the \A manifold predissociates {with a}
$\sim5\times10^{-7}$ {probability}
due to spin-orbit mixing with the $B$ state.
The measured values of \Ppd {imply a}
$\sim50\%$ {predissociative molecule loss}
after scattering $10^4$ photons, {suggesting}
that a MOT of CaH is feasible. We further extract the dipole matrix elements for all transitions connecting the ground \X$(v'')$ states to the excited \B$(v')$ states. This allows us to predict a viable stimulated Raman adiabatic passage (STIRAP) pathway to controllably dissociate the CaH molecules and subsequently trap the {resulting} ultracold hydrogen atoms{, which is a prospective goal for molecular laser cooling and cold chemistry research \cite{Lane_UltrcoldHydrogen_2015}}.      

\section{Calculation of molecular potential energies}
\label{sec:PES}
The starting point for our calculations is the construction of the potential energy surface (PES) for CaH. All calculations are performed using the Molpro program \cite{molpro,werner2012molpro,werner2020molpro}.  We adopt a
basis set and active space as {in Ref.}
\cite{cah2017}, where we use cc-pwCVQZ \cite{ccpwCVQZ} for the Ca atom and aug-cc-pVQZ \cite{ccpvqz} for the H atom.  Calculations are performed in $C_{2v}$ symmetry, which is the nearest Abelian point group to $C_{\infty v}$.  Orbitals are generated with a restricted Hartree-Fock (RHF) formalism, then further optimized in a state-averaged complete active space self-consistent field (SA-CASSCF) \cite{casscf} calculation involving 3 active electrons and 9 active orbitals.  For the $\Sigma^{+}$ states, 4 states are averaged at equal weights in the SA-CASSCF calculation, with (5,2,2,0) closed and (9,4,4,1) occupied orbitals. 

For the $A^2\Pi$ state, since only Abelian group symmetries are available, a two-state SA-CASSCF calculation with the same active space is performed in $C_{2v}$ symmetry involving symmetries 2 and 3 of equal weight to represent the $C_{\infty v}~A^2\Pi$ state.  These wavefunctions are then used in a multireference configuration interaction calculation with Davidson corrections for higher excitations (MRCI+Q) \cite{mrci1988,MRCI1992,mrci2011}.  Here, (3,1,1,0) orbitals make up the core, (5,2,2,0) are closed and (9,4,4,1) are occupied.  Electron correlation involving double and single excitations is allowed. The spin-orbit {interaction} is incorporated at the MRCI level using the Breit-Pauli Hamiltonian \cite{SOC2000}.

\section{Calculation of Franck-Condon factors}
Next, we employ the vibrational wavefunctions obtained in Section \ref{sec:PES} to calculate the Frank-Condon factors (FCFs) for the CaH transitions of interest. FCFs are calculated using a grid representation of the vibrational wavefunctions.  A spline interpolation is fit to the potential energy surfaces calculated in Molpro to create the potential energy functions, $V(r)$.  The real space kinetic energy operator is approximated with the Colbert-Miller derivative \cite{colbertmiller1992}. Nonadiabatic coupling vectors are computed analytically with the CP-MCSCF program \cite{nacme1991analytical} in Molpro and fit to a spline interpolation.  They are incorporated into the Hamiltonian by directly adding the nonadiabatic coupling to the momentum operator \cite{roibaer}. The Hamiltonian is diagonalized to obtain eigenvalues and eigenvectors.  Our calculations converge with a grid-spacing $(dr)$ of 0.007 $a_0$ and a box size of 16.5 $a_0$. Details are discussed in Appendix \ref{sec:nacme}.

\begin{table}[]
\centering
\begin{tabular}{|c|c|c|c|}
\hline
Transition & \begin{tabular}[c]{@{}c@{}}Vibrational Quanta\\ $(v'')$\end{tabular} & \begin{tabular}[c]{@{}c@{}}FCF Calculated \\ $(f_{0v''})$\end{tabular} & \begin{tabular}[c]{@{}c@{}}FCF Measured\\ $(f_{0v''})$\end{tabular} \\ \hline
\multirow{4}{*}{$A\rightarrow X$} & 0 & 0.9788 & 0.9572(43) \\ \cline{2-4} 
 & 1 & 0.0205 & 0.0386(32) \\ \cline{2-4} 
 & 2 & 6.8$\times 10^{-4}$ & 4.2(3.2)$\times 10^{-3}$ \\ \cline{2-4} 
 & 3 & 4.1$\times 10^{-5}$ & - \\ \hline
\multirow{4}{*}{$B\rightarrow X$} & 0 & 0.9789 & 0.9807(13) \\ \cline{2-4} 
 & 1 & 0.0192 & 0.0173(13) \\ \cline{2-4} 
 & 2 & 1.8$\times 10^{-3}$ & 2.0(0.3)$\times10^{-3}$ \\ \cline{2-4} 
 & 3 & 1.4$\times 10^{-4}$ & - \\ \hline
\end{tabular}
\caption{{The} calculated
Franck-Condon factors (FCFs) for CaH{, compared to}
experimental FCFs
\cite{Vasquez-Carson_2022_CaH}.
{The experimental FCFs are derived from measured vibrational branching ratios}. Note that the active space is optimized for the $B$ state {in}
this work.
}
\label{tab:VBRs}
\end{table}

We compare our calculated FCFs to previous experimental measurements
\cite{Vasquez-Carson_2022_CaH} in Table \ref{tab:VBRs}. We choose the active space which matched \B and \X state FCFs and vibrational energies in all calculations, since MRCI spin-orbit coupling (SOC) requires the same active space for all involved states. Therefore, the FCFs for $A^2\Pi$ could be improved with varied active space, but a compromise is made to estimate SOC splittings.  Despite this compromise, we find the \A potential {has}
the correct shape but a slightly incorrect equilibrium bond length. More details are in
{Appendix \ref{sec:nacme}.}

\section{\B Predissociation estimate}
\label{sec:prediss_calc}

Predissociation {probability} estimates are computed using an optical absorbing potential {with}
previously predicted scattering cross sections close to experiment \cite{neuhauser1989application,neuhauser1990state,neuhauser1995absorber}. {An absorbing
potential resembling a decaying half-parabola of the form $-iV(r-r_0)^{2}/w^{2}$ is added to the \X potential energy starting and centered at $r_0$ = 8 $a_0$ with a width $w =$ 8 $a_0$ and a depth of $V=$ 0.2 a.u. (4.4$\times 10^4$~cm$^{-1}$). Results are insensitive to absorber placement as long as it is placed along the potential energy surface's asymptote \cite{neuhauser1995absorber} and has a width larger than the typical de Broglie wavelength \cite{Neuhauser_1989_absorber}}.  This creates a channeled flux equation which imposes a boundary condition on the wavefunction and eigenvalues {attain}
an imaginary component. {Details can be found in Appendix \ref{sec:nacme}.} 

This component, {such as}
the imaginary eigenvalue of $B(v'=0)$, is directly related to the nonadiabatic coupling between that vibrational wavefunction and the $X$ continuum (where we place the absorber) as the nonradiative transition rate $A_{\mathrm{NR}}$.
We estimate {the} predissociation {probability} as the ratio of the calculated nonradiative ($A_{\mathrm{NR}}$) and radiative ($A_{R}$) decay rates,
$A_{\mathrm{NR}}/(A_{\mathrm{NR}}+A_{R}).$

\section{\B Predissociation measurement}
\label{sec:measurement}
\subsection{Experimental setup}
\label{subsec:setup}

The experimental setup has been {previously} described
\cite{Vasquez-Carson_2022_CaH}. Briefly, CaH is generated through ablation of a CaH$_2$ target by a pulsed Nd:YAG laser at a $\sim1$~Hz {rate}. CaH is buffer-gas cooled by helium at 6~K and ejected from the cell aperture to form a beam. The molecules are predominantly in the \X($v''=0$) state. The beam of CaH then enters a high-vacuum chamber which
is divided into four regions: state preparation, interaction, cleanup, and detection,
as shown in Fig. \ref{fig:CaH_PES}(b). 
In the first three regions, the molecular beam intersects with transverse lasers that address $X \rightarrow A$ or $X \rightarrow B$ transitions. These lasers can be switched on and off by independent optical shutters. The laser beams are multipassed to increase the interaction time with the molecular beam. In the detection region, we apply a single-pass $X \rightarrow A$ or $X \rightarrow B$ light and use an iXon888 {electron multiplying charge coupled device (EMCCD) camera}
and {a} Hamamatsu R13456 {photomultiplier tube}
to collect the laser-induced fluorescence (LIF) signals for spatially and temporally {resolved} detection.
Every molecule scatters
$\sim20$ photons in the detection region, which
implies that we are not sensitive to the initial
spin-rotation and hyperfine
distribution.
All {addressed} transitions are from the \X ($N''=1$) state ($N$ is the rotational quantum number) to \A ($J'=1/2$) ($J$ is the total angular momentum quantum number) or \B ($N'=0$) states in order to obtain rotational closure \cite{DiRosaEPJD04_LaserCoolingMolecules}. We use  electro-optic modulators (EOMs) to generate sidebands on all lasers to cover all hyperfine states (HFS) as well as to address spin-rotation manifolds. {The}
transitions used here are first measured experimentally with HFS resolution. Details of the lasers and transition frequencies can be found in Appendix \ref{sec:laser_conditions}.

To concisely describe the lasers used in this study we adopt the notation $M^{{R}}_{v'-v''}$, which denotes the transitions addressed and the spatial positions of the lasers. $M$ {is}
$A$ or $B$, representing the electronic state of the excited manifold. ${R}$ 
{is}
$S$, $I$, $C$, or $D$
{(state preparation, interaction, cleanup, or detection region)}. In addition, {the}
$F_{Mv'v''}$ notation
describes the vibrational branching ratios (VBRs) from either \A or \B states (represented by $M$) to \X states. For example, $F_{B01}$ is the VBR from \B $(v'=0)$ to \X $(v''=1)$. We
use similar notation, $F_{B0a}$ and $F_{B1a}$, to represent predissociation probabilities from \B $(v'=0)$ and $(v'=1)$ states.

\begin{figure}[ht!]
   \centering
   \includegraphics[scale=0.5]{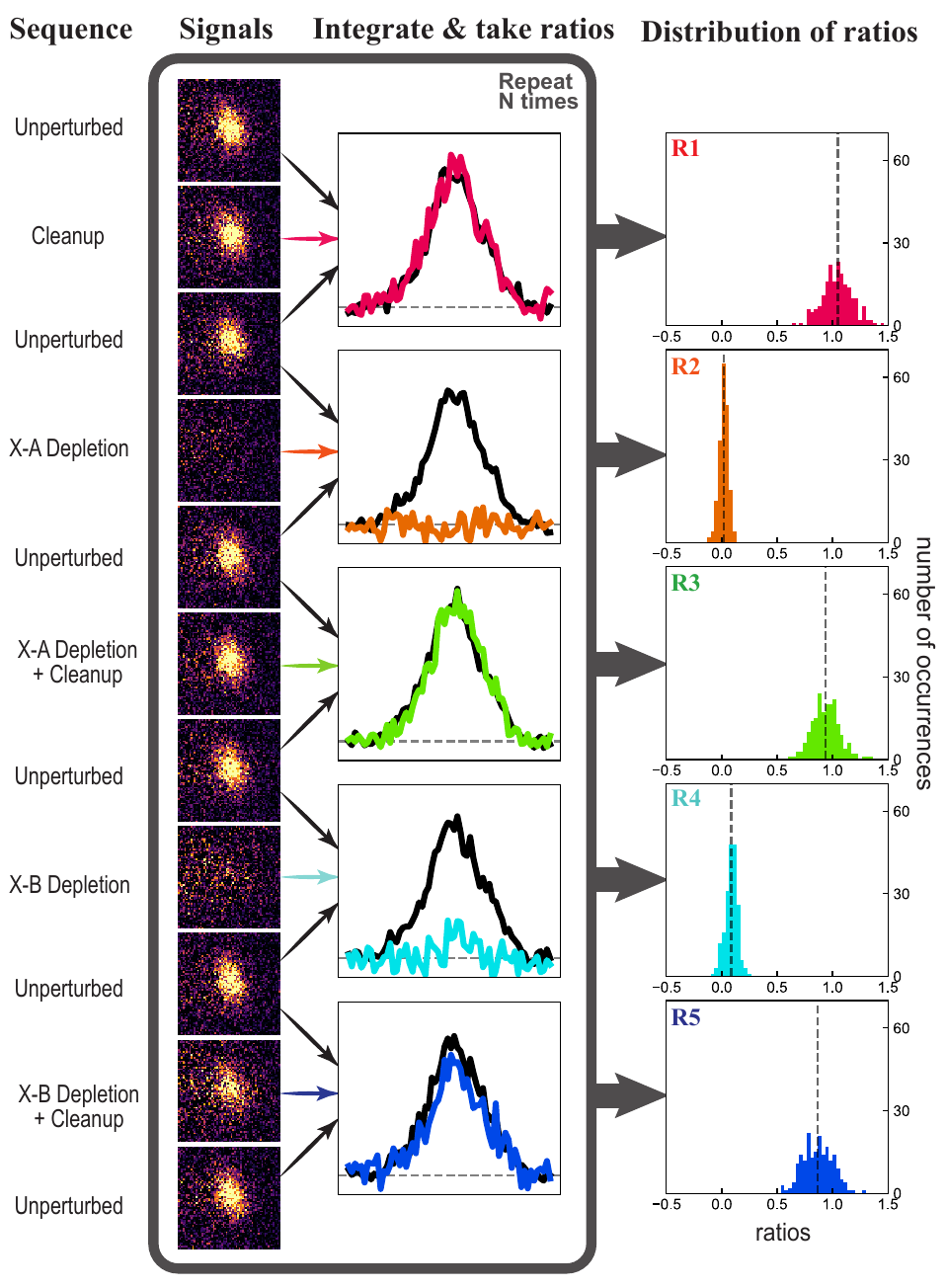}    \caption{
   Illustration of the ratio extraction process
   {for the} \B($v'=0$) predissociation measurement. We run the
   stages sequentially with a{n interlaced} reference stage,
   and collect LIF with an EMCCD. We integrate the images along both axes to obtain the signals, which we then used to calculate ratios. By repeating the {entire}
   sequence N times, we collect N sets of five ratios. {Here}
   we first show examples
   of one-shot camera images.
   We {then} present the integrated signal along one axis, {using}
   colored traces for science stages and black
   for reference stages (horizontal lines {are}
   the baselines).
   {Finally} we show the histograms of the five ratios.
   Vertical dashed lines represent the means of the ratios. 
   }
   \label{fig:graphical_ratios}
\end{figure}

\subsection{\B$(v' = 0)$ predissociation measurement method}
\label{subsec:v0_measurement}

To measure the predissociation probability of the \B($v'=0$) state,
we need to scatter
many photons
via \B($v'=0$)
and detect population loss that cannot be explained by known effects, predominantly rovibrational losses. To characterize the loss
we design several {experimental} stages, each {stage corresponding to a unique configuration of lasers interacting with the molecular beam}.
We monitor the population of the $v''=0$ ground state in the detection region by detecting
LIF signals from the $B^D_{0-0}$ laser. For this measurement we employ
{6} stages. By defining temporally stable parameters that describe the properties of our system, we can express the molecular population distribution at {each}
stage.

For example, in the \textit{Unperturbed} stage we detect $X(v''=0)$ population
denoted {by}
$N$. This is
the calibration {signal used as a reference.}
In the \textit{Cleanup} stage we apply the $B^{C}_{0-1}$ laser, and the {resulting} $X(v''=0)$ population is $N + n_1 N \kappa F_{B00} / \mathcal{F}_{B0}$ where $n_1$ is the normalized
natural population {of $X(v''=1)$}, $\kappa$ is the cleanup {laser} efficiency,
and {$\mathcal{F}_{B_0} \equiv F_{B0a} + \sum_{i\neq1}F_{B0i}$}
is the VBR normalization factor. This factor accounts for the discrete probability distribution of decay processes based on the VBRs and \Ppd. By taking the ratio of the integrated signal of the $X(v''=0)$ population from the \textit{Cleanup} stage with signal from the \textit{Unperturbed} stage, we get the parametrized ratio $R_1 = 1 + n_1 \kappa F_{B00} / \mathcal{F}_{B0}$. In addition to the \textit{Unperturbed} and \textit{Cleanup} stages, we have four more stages in this measurement, resulting in a total of 5 ratios and 5 parameters (including \Ppd). The details of all the stages, such as the laser configurations and expressions for the normalized signal,
{are} in Table \ref{tab:v0 table} and {Appendix \ref{sec:append_stages}}.
{Thus} we acquire 5 equations ({measured} ratios equal to the parametrized expressions) and 5 variables. 
We can solve the
equations and express $F_{B0a}$ via $R_i$s. By {precisely} measuring $R_i$ we can estimate the \B $(v'=0)$
predissociation probability. 

\begin{table*}[htbp]
\centering
\begin{tabular}{|c|c|c|c|}
\hline
Purpose       & Upstream Laser Config         & {Downstream Normalized \X $(v''=0)$ State Pop}                                                        & Averaged Signal Ratio \\ \hline
Unperturbed   & -                             & 1                                                                                       & -                     \\ \hline
Cleanup  & $B^{C}_{0-1}$                 & $1 + n_1 \kappa F_{B00}/\mathcal{F}_{B_0}$                                              & 1.05(2)               \\ \hline
X-A Cycling & $A^{I}_{0-0}$                 & $d_A$                                                                                   & 0.018(6)              \\ \hline
X-A Cycling + Cleanup   & $A^{I}_{0-0}  +   B^{C}_{0-1}$ & $d_A + [(1-d_A) F_{A01} / \mathcal{F}_{A_0} + n_1] \kappa F_{B00} / \mathcal{F}_{B_0}$  & 0.94(2)              \\ \hline
X-B Cycling & $B^{I}_{0-0}$                 & $d_B$                                                                                   & 0.086(8)              \\ \hline
X-B Cycling + Cleanup   & $B^{I}_{0-0}  +  B^{C}_{0-1}$  & $d_B + [(1-d_B) F_{B01} / \mathcal{F}_{B_1} + n_1 ] \kappa F_{B00} / \mathcal{F}_{B_0}$ & 0.87(2)              \\ \hline
\end{tabular}
\caption{Experimental stages for $B(v'=0)$ state predissociation measurement. In the second column,
$M^{{R}}_{v'-v''}$ {denotes}
the laser information. $M$ {is}
$A$ or $B$, representing the electronic excited state. ${R}$ 
{denotes the region}
$S$, $I$, or $C$ {(see text).}
The third column {contains}
the normalized ground-state
populations
using unknown variables and calculated VBRs.
{The} five variables
$n_1$, $\kappa$, $F_{B0a}$, $d_A$ and $d_B$ represent $X(v''=1)$ state natural population, cleanup efficiency of laser $B^{C}_{1-0}$, $B(v'=0)$ state predissociation probability, depletion efficiency of laser $A^{I}_{0-0}$ and depletion efficiency of laser $B^{I}_{0-0}$.
We denote the VBR normalization factors as {$\mathcal{F}_{A_0} \equiv \sum_{i\neq 0}F_{A0i}$, $\mathcal{F}_{B_0} \equiv F_{B0a} + \sum_{i\neq1}F_{B0i}$, and $\mathcal{F}_{B_1} \equiv F_{B0a} + \sum_{i\neq 0}F_{B0i}$.} 
{Additional information is} in Appendix {\ref{sec:append_stages}}.
}
\label{tab:v0 table}
\end{table*}

\subsection{\B $(v'=1)$ predissociation measurement method}
\label{subsec:v1_measurement}
For the $B(v'=1)$ state,
predissociation is {also} measured within the
framework of stages. We {implement}
two different methods, each consisting of multiple laser configurations,
to measure the same quantity. In method I we use 6 stages,
always monitoring the $X(v''=0)$ population downstream using laser $A^{D}_{0-0}$. The aim is to populate  $X(v''=1)$
via an off-diagonal pumping laser $A^{S}_{1-0}$ and perform {optical}
cycling between $X(v''=1)$ and  $B(v'=1)$.
We expect to see an increase of the $X(v''=0)$ population {as a result of the}
cycling. We
repump the molecules
remaining in $X(v''=1)$
to $v''=0$. {The recovered}
population
might {be less than}
expected {due to}
vibrational loss. By ruling out other effects,
we
attribute the loss to $B(v'=1)$ predissociation. The details of the 6 stages {are}
in Table \ref{tab:v1 table 1}.

Method II
differs
in {several}
ways. We monitor the $X(v''=1)$ population instead of $v''=0$,
accounting for loss to both $v''=0$ and $v''=2$
with {a sufficient}
signal-to-noise ratio (SNR) using laser $B^{D}_{1-1}$.
{The} 10 stages in this method lead to 9 measured ratios. And {the 7 required}
parameters {imply that there are}
more equations than variables. To find the optimal solution {of}
this over-constrained system, we define a least-squares objective function and use the Levenberg-Marquardt algorithm to search for the local minimum in the parameter space with reasonable initial guesses.

\begin{figure}[ht]
   \centering
   \includegraphics[scale=0.7]{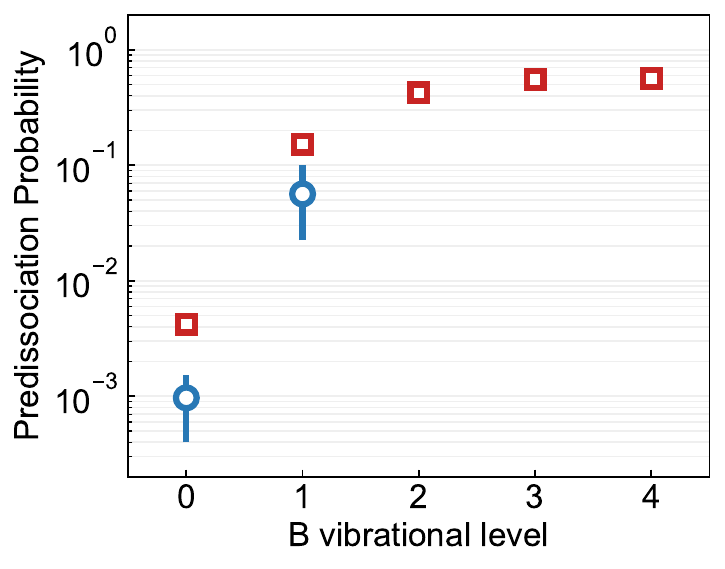}
   \caption{
   {CaH p}redissociation.
   Red squares are theoretical
   results {for}
   nonradiative decay rates {of}
   different vibrational states of \B. Blue circles are experimental results, where error bars represent the 95\% confidence interval.}
   \label{fig:Prediss_fig}
\end{figure}

\begin{table*}[htbp]
\centering
\begin{tabular}{|c|c|c|c|}
\hline
Purpose                                                                                 & Upstream Laser Config                          & {Downstream Normalized \X $(v''=0)$ State Pop}                                                                                                                                            & Averaged Signal Ratio \\ \hline
Unperturbed                                                                             & -                                              & 1                                                                                                                                                                           & -                     \\ \hline
State Prep                                                                              & $A^{S}_{1-0}$                                 & $1 - a$                                                                                                                                                                     & 0.22(2)               \\ \hline
Cleanup                                                                                 & $A^{C}_{0-1}$                                  & $1 + n_1 \kappa F_{A00}/\mathcal{F}_{A_1}$                                                                                                                                  & 1.10(3)             \\ \hline
State Prep + Cleanup                                                                 & $A^{S}_{1-0}  +  A^{C}_{0-1}$                  & $1 - a + (n_1 + a F_{A11} / \mathcal{F}_{A_2}) \kappa F_{A00} / \mathcal{F}_{A_1}$                                                                                          & 1.01(3)             \\ \hline
State Prep + X-B 1-1 Cycling                                                         & $A^{S}_{1-0}  +   B^{I}_{1-1}$                & $1 - a + (n_1 + a F_{A11} / \mathcal{F}_{A_2}) d_B F_{B10} / \mathcal{F}_{B_2}$                                                                                             & 0.39(2)             \\ \hline
\begin{tabular}[c]{@{}c@{}}State Prep + \\ X-B 1-1 Cycling + Cleanup\end{tabular} & $A^{S}_{1-0}  +   B^{I}_{1-1}  +  A^{C}_{0-1}$ & \begin{tabular}[c]{@{}c@{}}$1 - a + (n_1 + a F_{A11} / \mathcal{F}_{A_2}) (d_B F_{B10} / \mathcal{F}_{B_2}$ \\ $+ (1 - d_B) \kappa F_{A00} / \mathcal{F}_{A_1})$\end{tabular} & 0.40(2)             \\ \hline
\end{tabular}
\caption{Method I of \B$(v'=1)$
predissociation measurement. Notation {is}
similar {to}
Table \ref{tab:v0 table}.
In the third column, the variables include $a$, $n_1$, $\kappa$, $F_{B1a}$ and $d_B$, representing state preparation (from $X(v''=0)$ to $X(v''=1)$) efficiency, $X(v''=1)$
natural population, cleanup efficiency of laser {$A^{C}_{0-1}$}, $B(v'=1)$
predissociation probability {and} depletion efficiency of laser $B^{I}_{1-1}$.
The VBR normalization factors {are}
{$\mathcal{F}_{A_1} \equiv \sum_{i\neq1}F_{A0i}$, $\mathcal{F}_{A_2} \equiv \sum_{i\neq0}F_{A1i}$, and $\mathcal{F}_{B_2} \equiv F_{B1a} + \sum_{i\neq1}F_{B1i}$.} 
Additional information is in Appendix {\ref{sec:append_stages}}.
}
\label{tab:v1 table 1}
\end{table*}

\begin{table*}[htbp]
\centering
\begin{tabular}{|c|c|c|c|}
\hline
Purpose                                   & Upstream Laser Config                         & {Downstream Normalized \X $(v''=1)$ State Pop}                                                                                                              & Avg Ratio \\ \hline
State Prep + Cleanup v0                   & $A^{S}_{1-0} + A^{C}_{1-0}$                 & $n_1 + (a + \kappa_1 - a \kappa_1) F_{A11}/\mathcal{F}_{A_2}$                                                                                 & -                     \\ \hline
Unperturbed                               & -                                             & $n_1$                                                                                                                                             & 0.13(3)               \\ \hline
State Prep                                & $A^{S}_{1-0}$                                & $n_1 + a F_{A11}/ \mathcal{F}_{A_2} \equiv \mathcal{Z}$                                                                                       & 0.89(4)               \\ \hline
Cleanup v0                                & $A^{C}_{1-0}$                                 & $n_1 + \kappa_1 F_{A11}/\mathcal{F}_{A_2}$                                                                                                      & 0.93(4)               \\ \hline
State Prep + X-A 1-1 Cycling              & $A^{S}_{1-0} + A^{I}_{1-1}$               & $\mathcal{Z} (1 - d_A)$                                                                                                                       & 0.03(3)               \\ \hline
State Prep + X-A 1-1 Cycling + Cleanup v0 & $A^{S}_{1-0} +  A^{I}_{1-1} + A^{C}_{1-0}$ & $\mathcal{Z} (1 - d_A) + (1 - a + \mathcal{Z} d_A F_{A10} / \mathcal{F}_{A_3}) \kappa_1 F_{A11} / \mathcal{F}_{A_2}$                          & 0.33(3)               \\ \hline
State Prep + X-A 1-1 Cycling + Cleanup v2 & $A^{S}_{1-0} +  A^{I}_{1-1} + A^{C}_{1-2}$ & $\mathcal{Z} (1 - d_A) + (a F_{A12} / \mathcal{F}_{A_2} +  \mathcal{Z} d_A F_{A12} / \mathcal{F}_{A_3}) \kappa_2 F_{A11} / \mathcal{F}_{A_4}$ & 0.57(4)               \\ \hline
State Prep + X-B 1-1 Cycling              & $A^{S}_{1-0} + B^{I}_{1-1}$               & $\mathcal{Z} (1 - d_B)$                                                                                                                       & 0.12(3)             \\ \hline
State Prep + X-B 1-1 Cycling + Cleanup v0 & $A^{S}_{1-0} +  B^{I}_{1-1} + A^{C}_{1-0}$ & $\mathcal{Z} (1 - d_B) + (1 - a + \mathcal{Z} d_B F_{B10} / \mathcal{F}_{B_2}) \kappa_1 F_{A11} / \mathcal{F}_{A_2}$                          & 0.35(3)               \\ \hline
State Prep + X-B 1-1 Cycling + Cleanup v2 & $A^{S}_{1-0} +  B^{I}_{1-1} + A^{C}_{1-2}$ & $\mathcal{Z} (1 - d_B) + (a F_{A12} / \mathcal{F}_{A_2} +  \mathcal{Z} d_B F_{B12} / \mathcal{F}_{B_2}) \kappa_2 F_{A11} / \mathcal{F}_{A_4}$ & 0.42(3)               \\ \hline
\end{tabular}
\caption{Method II of \B$(v'=1)$
predissociation measurement. In the third column, {the 7}
variables
include $a$, $n_1$, $\kappa_1$, $\kappa_2$, $F_{B1a}$, $d_A$ and $d_B$, representing state preparation (from $X(v''=0)$ to $X(v''=1)$) efficiency, $X(v''=1)$
natural population, cleanup efficiency of laser $A^{C}_{1-0}$,
cleanup efficiency of laser $A^{C}_{1-2}$, $B(v'=1)$
predissociation probability, depletion efficiency of laser $A^{I}_{1-1}$ {and} depletion efficiency of laser $B^{I}_{1-1}$.
The VBR normalization factors {are}
{$\mathcal{F}_{A_2} \equiv \sum_{i\neq0}F_{A1i}$, $\mathcal{F}_{A_3} \equiv \sum_{i\neq 1}F_{A1i}$, $\mathcal{F}_{A_4} \equiv \sum_{i\neq 2}F_{A1i}$, and $\mathcal{F}_{B_2} \equiv F_{B1a} + \sum_{i\neq1}F_{B1i}$.}
Additional information is in Appendix {\ref{sec:append_stages}}.
}
\label{tab:v1 table 2}
\end{table*}

\subsection{Predissociation measurement analysis}
\label{subsec:result_analysis}

The yield of our CBGB source {exhibits some slow}
drift.
In order to reduce
errors due to {molecule} number fluctuations, we {insert}
a reference stage before and after every other stage within a group when taking data. For example, in the $B(v'=0)$ predissociation measurement, data are taken in the following order: \textit{Unperturbed} $\rightarrow$ \textit{Cleanup} $\rightarrow$ \textit{Unperturbed} $\rightarrow$ \textit{X-A Cycling} $\rightarrow$ \textit{Unperturbed} ... \textit{X-B Cycling + Cleanup} $\rightarrow$ \textit{Unperturbed}. The reference stage for $B(v'=1)$ method I is \textit{Unperturbed}, while for method II it is \textit{State Prep + Cleanup v0}. To calculate the ratios, we divide the signal by the average signal from the calibration shots before and after. {The entire group}
of measurements {is repeated} multiple times.
The averaged ratios can be found in Tables \ref{tab:v0 table}, \ref{tab:v1 table 1}, and \ref{tab:v1 table 2}. The values in parentheses denote the $2\sigma$ statistical errors. A graphical representation of the analysis process and histograms of all five ratios can be found in Fig. \ref{fig:graphical_ratios}.

With the ratios measured, we use a bootstrap method \cite{Efron_1979_bootstrapping,EfroTibs_1993_bootstrapping,davison_hinkley_1997_bootstrap} to derive the mean values and build the confidence intervals of the predissociation probabilities depicted in Fig. \ref{fig:Prediss_fig}. This method is particularly useful as it does not require any
assumptions about the data such as independence assumptions typically made for standard error calculations. We consider several other analysis methods, such as pairwise bootstrapping and regular error propagation, {and}
the outcomes are all {in agreement with}
each other. Details of the bootstrap method {are}
in Appendix \ref{sec:bootstrapping}. 

After
considering all systematic effects and analyzing statistical errors, we {find}
the predissociation probability for the \B$(v'=0)$ state to be $0.00097^{+0.00059}_{-0.00057}$ and for the \B$(v'=1)$ state to be $0.056^{+0.044}_{-0.034}$. The reported value for $B(v'=1)$ is the average of the two methods (method I yields $0.079^{+0.021}_{-0.017}$ and method II yields $0.033^{+0.013}_{-0.011}$), and the 95\% confidence interval is the largest of the two methods combined.  These values are consistent with the probabilities calculated in Sec. \ref{sec:prediss_calc} within the order of magnitude. {Other potential loss channels are discussed in Appendix \ref{sec:loss}.}  In addition, {to demonstrate the robustness of our measurements to small variations in FCFs,} we perform a comparative analysis by utilizing the FCFs obtained {in}
previous theoretical work on CaH \cite{Gao_Summary_2014,Ramanaiah_1982_CaH_FCF}.
The results {consistently produce nonzero predissociation probabilities and} are within error bars of each other. The {sharp}
monotonic increase in \Ppd seen in Fig. \ref{fig:Prediss_fig} can be understood {as}
a bound molecul{e}
quantum tunneling through the \B potential
into the \X continuum
at the same energy. As the energy of the incident quantum state increases, {so does} the transmission probability{, which is aided by stronger wavefunction overlaps}.

\begin{table*}[htpb]
\begin{tabular}{c|ccc}
 & \begin{tabular}[c]{@{}c@{}}$A^2\Pi_x$\\ $(v=0,m_s=1/2)$\end{tabular} & \begin{tabular}[c]{@{}c@{}}$A^2\Pi_y$\\ $(v=0,m_s=1/2)$\end{tabular} & \begin{tabular}[c]{@{}c@{}}$B^2\Sigma^+$\\ $(v=0,m_s=-1/2)$\end{tabular} \\ \hline
\begin{tabular}[c]{@{}c@{}}$A^2\Pi_x$\\ $(v=0,m_s=1/2)$\end{tabular} & 0 & -35.5$i$ & 21.5 \\
\begin{tabular}[c]{@{}c@{}}$A^2\Pi_y$\\ $(v=0,m_s=1/2)$\end{tabular} & 35.5$i$ & 0 & -21.5$i$ \\
\begin{tabular}[c]{@{}c@{}}$B^2\Sigma^+$\\ $(v=0,m_s=-1/2)$\end{tabular} & 21.5 & 21.5$i$ & 1400
\end{tabular}

\bigskip

\begin{tabular}{c|cccc}
 & \begin{tabular}[c]{@{}c@{}}$A^2\Pi_x$\\ $(v=1,m_s=1/2)$\end{tabular} & \begin{tabular}[c]{@{}c@{}}$A^2\Pi_y$\\ $(v=1,m_s=1/2)$\end{tabular} & \begin{tabular}[c]{@{}c@{}}$B^2\Sigma^+$\\ $(v=0,m_s=-1/2)$\end{tabular} & \begin{tabular}[c]{@{}c@{}}$B^2\Sigma^+$\\ $(v=1,m_s=-1/2)$\end{tabular} \\ \hline
\begin{tabular}[c]{@{}c@{}}$A^2\Pi_x$\\ $(v=1,m_s=1/2)$\end{tabular} & 0 & -35.5$i$ & 21.5$f$ & 21.5 \\
\begin{tabular}[c]{@{}c@{}}$A^2\Pi_y$\\ $(v=1,m_s=1/2)$\end{tabular} & 35.5$i$ & 0 & -$(21.5f)i$ & -21.5$i$ \\
\begin{tabular}[c]{@{}c@{}}$B^2\Sigma^+$\\ $(v=0,m_s=-1/2)$\end{tabular} & 21.5$f$ & $(21.5f)i$ & 64 & 0 \\
\begin{tabular}[c]{@{}c@{}}$B^2\Sigma^+$\\ $(v=1,m_s=-1/2)$\end{tabular} & 21.5 & 21.5$i$ & 0 & 1310
\end{tabular}
\caption{Spin-orbit matrices accounting for
vibrational {mixing}
of the $A$ and $B$ states. The $\Pi_x$ and $\Pi_y$ basis states split under SOC to produce $\Pi_{1/2}$ and $\Pi_{3/2}$ states. {The top matrix is for \As$(v'=0)$ and \B$(v'=0)$, while the bottom one is for}
\As$(v'=1)$, \B$(v'=1)$ and \B$(v'=0)$. The Franck-Condon factor $f$ is introduced to account for the off-diagonal vibrational wavefunction overlap. The diagonal terms represent the energies of
unperturbed states.  All values are in cm$^{-1}$.}
\label{table:SOC matrix}
\end{table*}

\section{\A Predissociation estimate}
\label{sec:SOC}
The \As state in
CaH does not undergo predissociation via the process described for the \B state.
However, spin-orbit coupling can induce mixing between the $A$ and $B$ states, leading to {non-vanishing} predissociation of the \A spin-orbit state.
For a linear molecule, the $z$-component of total angular momentum, $J_z$, is a good quantum number. {Therefore}
the spin-orbit component \A
{can interact with} $B^2\Sigma^+(J_z=1/2)$ {due to the same $J_z$ value}.
A similar interaction exists between \A and $X^2\Sigma^+(J_z=1/2)$
but the energy separation is much larger ($\sim$14,000~cm$^{-1}$) compared to that between the $A$ and $B$ states ($\sim$1,400~cm$^{-1}$). Higher vibrational states of the $X$ manifold {are}
closer in energy to $A$
but the effective coupling to the states relevant for laser cooling is
weaker due to a poor vibrational wavefunction overlap.

We estimate the mixing between the $A(v'=0)$ and the $B(v'=0)$ states separated by 1400~cm$^{-1}$. The spin-orbit parameters are obtained with the Breit-Pauli Hamiltonian at the MRCI level \cite{SOC2000} and are given in Table \ref{table:SOC matrix}. Diagonalization of this Hamiltonian matrix leads to a 0.05\% $B(v'=0)$ {admixture}
into the $A(v'=0)$ state. Similarly, we can compute the mixing between $A(v'=1)$, $B(v'=0)$, and $B(v'=1)$.
The coupling between $A(v'=1)$ and $B(v'=1)$
is expected to be similar to the case of $v'=0$ since the energy difference of 1310~cm$^{-1}$ is similar to that in the case of $v'=0$. However, the $A(v'=1)$ and $B(v'=0)$ states are only 64~cm$^{-1}$ apart, hence {even} a small {FCF can}
lead to significant mixing. Note that the measured FCF for the $A(v'=0)\rightarrow X(v''=1)$ transition is 4\% (Table \ref{tab:VBRs}) and that {our calculated}
$A-B$ bond length difference is smaller than the $X-A$ bond length. We use
$f=5$\% as an upper limit for the $A(v'=1)\rightarrow B(v'=0)$ FCF. Diagonalization of the corresponding Hamiltonian matrix in Table \ref{table:SOC matrix} {yields}
a 8.4\% $B(v'=0)$ character {for}
$A(v'=1)$. Combining {these admixtures}
with the measured \Ppd for $B(v'=0,1)$, we estimate that the $A(v'=0)$ state
very weakly predissociates with a probability of $\sim5 \times 10^{-7}$ {and}
the $A(v'=1)$ state {with}
a higher probability of $\sim3\times 10^{-5}$. The {FCF}
used here is an upper-bound value and therefore the estimated probabilities serve as upper bounds.

\section{Controlled Dissociation Pathway}
{As mentioned in Sec. \ref{sec:Intro}, an enticing application of ultracold CaH and other molecules is controlled dissociation into fragments that are not directly laser-coolable, such as H.}
In order to {trap the resulting H atoms,}
the{ir} maximum kinetic energy
must be below typical optical trap depths. A magic-wavelength trap for H atoms at 513~nm has a depth of 1.2~kHz
per 10~kW/cm$^2$ \cite{Kawasaki_2015_magic_hydrogen_trap}.
{A practical dipole trap with an intensity of at most}
$\sim$100~kW/cm$^2$
{would result in} a maximum trap depth of only $\sim0.5~\mu$K.
{Since}
the binding energy of
\B$(v'=0)$ corresponds to a temperature of $\sim1,000$~K, {the trapping of the}
fragments relies on the ability to dissociate the molecules as closely {as possible} to the threshold
\cite{Lane_UltrcoldHydrogen_2015}{, such as} via a stimulated two-photon {process \cite{ZelevinskyMcGuyerNJP15_Sr2Spectroscopy,McDonald_2016_photodissociation}.}

Stimulated Raman adiabatic passage (STIRAP) is a technique that has been successfully employed to generate ground-state bialkali molecules starting from a weakly bound state \cite{Vitanov_2017_STIRAP,Moses_2017_polar_molecules}. Although STIRAP has been predominantly demonstrated for adiabatic population transfers from weakly bound to deeply bound molecular states, the mechanism can be extended to unbound {continuum} states
\cite{Vardi_99_STIRAP,Rangelov_2007_STIRAP}. A prerequisite for efficient transfer is the identification of an intermediate state that strongly couples to both initial and final states. Additionally, a desirable intermediate state would be connected via readily accessible laser wavelengths to the initial and final states.

\begin{figure*}[htpb!]
   \centering
   \includegraphics[scale=0.75]{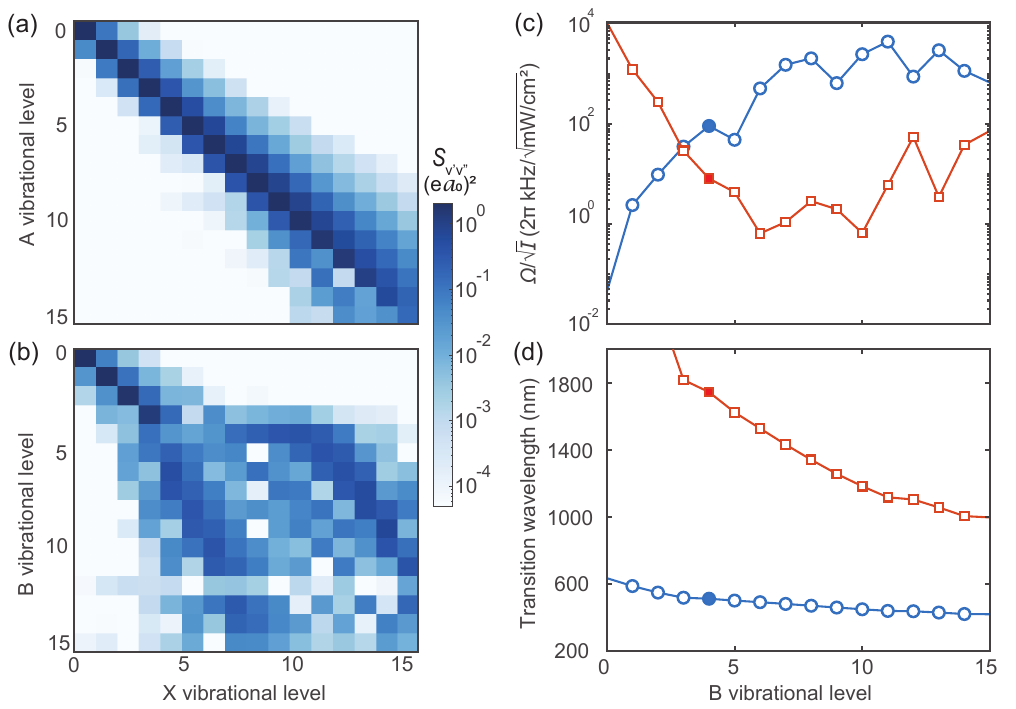}
   \caption{{Suggested} controlled dissociation pathway for CaH molecules. Line strengths ($S_{v'v''}$) in atomic units for
   dipole allowed transitions:
   (a) \X$(v'')\rightarrow$\A$(v')$ and (b) \X$(v'')\rightarrow$\B$(v')$. The $A$ state potential is more harmonic as is reflected by the diagonal $S_{v'v''}$. The $B$ state, however, {significantly} deviates from the diagonal starting around $v'=4$ because of
   the second potential minimum {at}
   $\sim6$~$a_0$ (Fig. \ref{fig:CaH_PES}(a)). Note that $B(v'=4)$ has comparable line strengths between $X(v''=0)$ and $X(v''=15)$. (c) Intensity-normalized Rabi rate ($\Omega/\sqrt{I} = \sqrt{S_{v'v''}}/\hbar$) for dipole transitions $X(v''=0)\rightarrow B(v')$ (red squares) and $B(v')\rightarrow X(v=15)$ (blue circles). Around $v'=4$ (shaded points), the Rabi rates {are}
   comparable.
   (d)
   Wavelengths in nanometers for optical transitions $X(v''=0)\rightarrow B(v')$ (blue circles) and $B(v')\rightarrow X$(continuum) (red squares). The theoretical energy differences are adjusted by a common offset of 240~\cm to match
   experimental data for $X(v''=0)\rightarrow B(v'=0,1,2)$ transitions
   \cite{Shayesteh_2013_CaH_fourier_spectra}. The wavelengths corresponding to $v'=4$ (shaded points) are 512.7 nm and 1744.7 nm.  
   }
   \label{fig:TDMs}
\end{figure*}

Molecular structure calculations give us access to branching ratios and line strengths for a multitude of vibrational levels, some of which
have {advantages}
for controlled dissociation. We calculate the dipole transition line strength $S_{v'v''}$, which is the square of the transition dipole moment $(|\bra{v'}\mu\ket{v''}|^2)$, for both \A$\rightarrow$\X and \B$\rightarrow$\X transitions (Figs. \ref{fig:TDMs}(a,b)). The PES for $A$ and $X$ state are similar in shape (Fig. \ref{fig:CaH_PES}(a)) which leads to highly diagonal transition strengths. However, the second minimum in the $B$ state PES leads to strong off-diagonal coupling starting around $v'=4$. This feature
enables
strong coupling {of}
$B(v'=4)$ {to both}
$X(v''=0)$ and $X(v''=15)$
(Fig. \ref{fig:TDMs}(c)). Our calculations do not show a significant coupling between the $B(v'=4)$ state and the vibrational{ly} excited states of the A manifold. {Here}
we
calculate the coupling {to the weakest bound state, rather than}
to the continuum,
for two reasons. First, we expect the coupling to the lowest-energy continuum state{s} and {to} the
{least-}bound state to be similar since their energy difference is only $\sim500$~cm$^{-1}$. Second, we expect the STIRAP process to be more efficient if all three states are bound states. Hence it is worthwhile to consider a transfer to $X(v''=15)$ {followed by}
photodissociation \cite{McDonald_2016_photodissociation} or Feshbach {dissociation}
\cite{Barbe_2018_RbSrFeshbachRes}.

{In Fig. \ref{fig:TDMs}(d) we} plot the laser wavelengths {required} to connect
$X(v''=0)$ {as well as}
the {ground-state} continuum to the $B$ manifold. We {estimate}
that the
{``upleg" } STIRAP
wavelength for $X(v''=0)\rightarrow B(v'=4)$ is 512.7~nm {while}
the {``downleg"} wavelength for {$B(v'=4)\rightarrow X(v''=v_{\mathrm{th}})$} is 1744.7~nm. Both these
wavelengths {are accessible via}
current technology such as Raman fiber amplifiers and difference-frequency generation (DFG). Thus we expect high-power and narrow-linewidth laser {source}s to be within reach for STIRAP.

\vspace{10pt}

\section{Conclusion}
Predissociation {is a}
challenge for laser cooling of {new}
molecular species. We have theoretically and experimentally studied {it for}
laser cooling CaH {as well as in the context of controlled ultracold dissociation.}
We find that the lowest-excited electronic state \A$(v'=0)$, which is the workhorse
for optical cycling, only weakly predissociates (\Ppd$\approx10^{-6}$) via spin-orbit coupling. The {next}
excited manifold \B, crucial for repumping vibrational dark states,
has much stronger predissociation by virtue of having the same symmetry as \X. We measure \Ppd for $B(v'=0)$ and $B(v'=1)$ states and obtain $\sim10^{-3}$ and $\sim6\times 10^{-2}$, respectively. This {sharp}
increase is substantiated by theoretical calculations, and we expect \Ppd$\rightarrow 1$ for $v'\gtrsim 4$.  {The results are summarized in Table \ref{tab:Ppds}.}

To obtain high photon scattering rates, one must repump the {$A(v'=0)\rightarrow X(v''=1)$} vibrational loss channel via the $B(v'=0)$ state. {Due to predissociation,}
we find that the optimal laser cooling scheme requires avoiding the $B(v'\ge 1)$ states {in favor of using}
the $A$ manifold. On
average, every {cycling} molecule will scatter $\sim20$ photons $(1/(1-F_{A00}))$ before being lost to $X(v''=1)$. Each of these molecules
only need{s} to scatter one photon via $B(v'=0)$ to return to cycling, but {it}
will {predissociate}
with a 0.1\% probability.
Hence {we} estimate that $\sim50\%$ of molecules will be lost to predissociation after scattering {the requisite}
$\sim10^4$ photons.

Last, we propose {to}
take advantage of the high predissociation probability for $B(v'=4)$ state to engineer a two-photon STIRAP pathway for transferring the molecular population from the ground $X(v''=0)$ state to the low-energy continuum $X(v''=v_{\mathrm{th}})$.
We find that
$B(v'=4)$
couples strongly to both {these} $X$ states via optical transitions at wavelengths within accessible laser technologies.

\begin{table*}[htbp]
\centering
\begin{tabular}{|c|c|c|c|c|c|c|}
\hline
State & \begin{tabular}[c]{@{}c@{}}Vibrational \\ quantum $(v'')$\end{tabular} & \begin{tabular}[c]{@{}c@{}}Radiative\\ lifetime (ns)\end{tabular} & \begin{tabular}[c]{@{}c@{}}Radiative decay\\ rate ($A_R$, s$^{-1}$)\end{tabular} & \begin{tabular}[c]{@{}c@{}}Nonradiative decay\\ rate ($A_{\mathrm{NR}}$, s$^{-1}$)\end{tabular} & \begin{tabular}[c]{@{}c@{}}Predissociation\\ (PD) probability\end{tabular} & \begin{tabular}[c]{@{}c@{}}Experimental\\ PD probability\end{tabular} \\ \hline
\multirow{7}{*}{B} & 0 & 52.0 & 1.924$\times10^{7}$& 8.040$\times10^{4}$ & 0.0042 & $0.00097_{-0.00057}^{+0.00059}$ \\ \cline{2-7} 
 & 1 & 54.3 & 1.842$\times10^{7}$ & 3.304$\times10^{6}$ & 0.1521 & $0.056^{+0.044}_{-0.034}$ \\ \cline{2-7} 
 & 2 & 58.9 & 1.698$\times10^{7}$ & 1.245$\times10^{7}$ & 0.4230 & - \\ \cline{2-7} 
 & 3 & 78.2 & 1.278$\times10^{7}$ & 1.571$\times10^{7}$ & 0.5514 & - \\ \cline{2-7} 
 & 4 & 59.2 & 1.688$\times10^{7}$ & 2.181$\times10^{7}$ & 0.5637 & - \\ \cline{2-7} 
 & 5 & 83.9 & 1.193$\times10^{7}$ & 5.482$\times10^{7}$ & 0.8213 & - \\ \cline{2-7} 
 & 6 & 84.4 & 1.185$\times10^{7}$ & 5.960$\times10^{7}$ & 0.8342 & - \\ \hline
\multirow{2}{*}{A} & 0 & 34.3 & 2.913$\times10^{7}$ & - & - & 5$\times10^{-7}$ \\ \cline{2-7} 
 & 1 & 34.5 & 2.902$\times10^{7}$ & - & - & 3$\times10^{-5}$ \\ \hline
\end{tabular}
\caption{Theoretical and experimental values of predissociation probability {for \B and \A}. Both radiative ($A_R$) and non-radiative ($A_{\mathrm{NR}}$) decay rates are calculated.
The radiative lifetime is
$\tau\equiv1/A_R$. Predissociation \Ppd probability is defined as the ratio of the nonradiative decay rate to the total decay rate $(A_R+A_{\mathrm{NR}})$. {Measurements}
of \Ppd are only provided for the \B$(v'=0)$ and $(v'=1)$ states. For the latter,
we report the mean of two different measurement methods described in Section \ref{subsec:v1_measurement}. The values given for the \A$(v'=0)$ and $(v'=1)$ states are estimated by calculating the spin-orbit mixing between
$A$ and $B$
as described in Section \ref{sec:SOC}.}
\label{tab:Ppds}
\end{table*}

%

\section*{Acknowledgements}
We thank Ian Lane for fruitful discussions of dissociation and Roi Baer for discussions of nonadiabatic couplings. {We thank Ye Tian for his assistance in implementing the bootstrapping method used in this study.} This work was supported by the ONR grant N00014-21-1-2644, AFOSR MURI grant FA9550-21-1-0069, and we acknowledge generous support by the W. M. Keck Foundation and the Brown Science Foundation. D.N. would like to acknowledge support from the BSF grant 2018368 and NSF-CHE grant 1763176.  A.N.A. acknowledges support from the NSF Center for Chemical Innovation Phase I grant CHE-2221453.  C.E.D. acknowledges support from NSF grant DGE-2034835.  

\appendix

\section{Other possible loss channels and their contributions}
\label{sec:loss}

Other {potential loss channels that disrupt optical cycling could}
lead to overestimating the predissociation probability. {The theoretical results are agnostic to such processes.}
We consider the following processes {that may contribute to population loss:}

\begin{itemize}
    \item Off-resonant excitation to the \B$(v'=0,N'=2)$ state. The nearest parity-allowed transition from $X(v''=0,N''=1)$
    is {to}
    $B(v'=0,N'=2)$
    which is
    768~GHz away from
    $B(v'=0,N'=1)$.
    The transition linewidth is $\sim2\pi \times 3$~MHz. Assuming a two-level-like system,
    the scattering rate is
    \cite{Metcalf_1999_laser_cooling_book}
    $$R_{\mathrm{sc}} = \frac{s\Gamma/2}{1 + s + (2\Delta/\Gamma)^2}.$$
    In our system, the saturation parameter {is} $s\lesssim1,000$, and thus
    $R_{\mathrm{sc}}\lesssim2$ s$^{-1}$. This rate is
    low compared to the estimated nonradiative decay rate of $10^5$~s$^{-1}$, {therefore}
    off-resonant excitation {should}
    not affect the result.

    \item
    External electric fields {$\epsilon$} can induce mixing between \B$(v'=0,N'=0)$ and $N'=1$ states ({e.g., Ref.}
    \cite{Norrgard_2016_thesis}, Section 8.4.2.1). For the \A state, the matrix element of the dipole operator $T^1_p (d)$ in Hund's case $a$ basis is $-\frac{1}{3}\epsilon d$. For the \B state expressed in Hund's case $b$ basis, we {first}
    project
    to Hund's case $a$ basis,
    then calculate the
    matrix element to be $-\frac{1}{2}\epsilon d$. The rotational spacing for \B
    is
    250~GHz, while the {$\Lambda$-doubling} for \A is
    26~GHz \cite{Shayesteh_2013_CaH_fourier_spectra}.  The {effective} decay rate is given by
    $$\frac{1}{\tau^{N=0,2}} = R_{\mathrm{sc}} \frac{(d \epsilon / 2)^2}{4 \omega^2 + \Gamma^2 / 4},$$
    where {for the $A$ and $B$ states respectively, the values are} $d$ = 2.57~D and 3.1~D,
    $\omega = 2\pi \times 13$~GHz
    and $2\pi \times 125$~GHz, {and}
    $\Gamma = 2\pi \times 5$~MHz
    and $2\pi \times 3$~MHz.
    {We assume} $R_{\mathrm{sc}}\approx1$~MHz. Since we electrically ground the entire vacuum chamber, the electric field amplitude inside the chamber should be
    $<100$~V$/$m. {We find that}
    the possible remixing rate is $6.2\times10^{-4}$~s$^{-1}$ for \A
    and $9.7\times 10^{-6}$~s$^{-1}$ for \B.
    These numbers are several orders of magnitude smaller than nonradiative decay rates {and should have a negligible effect}.

    \item Photon scattering along the {molecular} beam
    can cause acceleration or deceleration and affect signal strength. In the interaction region we scatter
    $<100$ photons per molecule. The laser beams are reflected in a zig-zag {pattern, i.e. the incidence}
    is not {perfectly perpendicular and there is a}
    projection {on the beam propagation}
    direction. The
    angle is estimated to be $\arctan(1/15) \approx 4^{\circ}$. {Hence}
    less than 10 photons worth of momentum is added to the molecular beam, and that would only {yield a}
    15~cm$/$s velocity change. The {beam} velocity
    is
    $\sim200$~m$/$s, implying that the {effect on the}
    signal {is at the}
    $8\times 10^{-4}$ {level} which {negligible.}

  \item {We consider o}ff-diagonal vibrational loss due to spin-orbit mixing. As discussed in Sec. \ref{sec:SOC}, the \B$(v'=0)$ state mixes with the \A$(v'=1)$ state at the 0.06\% level. This implies that population from 
  $B(v'=0)$
  can decay to $X(v''=1)$
  via
  $A(v'=1)$
  at a rate of $6\times 10^{-4} \times F_{A10} \approx 5\times 10^{-4}$. This value is 40 times smaller than the FCF {for the}
  $B(v'=0)\rightarrow X(v''=1)$ decay ($1.92\times 10^{-2}$) and hence is a negligible correction to our model. A similar argument holds for
  off-diagonal vibrational loss induced by spin-orbit mixing {of}
  the $B(v'=1)$ state with either
  $A(v'=0)$ or
  $A(v'=2)$.

\end{itemize}

\section{Laser parameters and spectroscopy of transitions used in this work}
\label{sec:laser_conditions}

Here we describe the
lasers {used}
in the experiment, and the transition frequencies.
All laser beams
pass through an {electro-optic modulator} (EOM) to generate
the sidebands needed to address HFS.

\begin{itemize}
\item In the state preparation region, {the}
$A^S_{1-0}$ light (637~nm)
is generated from two sets of injection-locked amplifiers (ILAs) to address the spin-rotation states, with 95~mW of power.

\item In the interaction region, multiple lasers are applied. $A^I_{0-0}$ (695~nm) or $A^I_{1-1}$ (693~nm) {light} is {derived} from two ILAs that provide 60~mW in total. $B^I_{0-0}$ (635~nm) or $B^I_{1-1}$ (636~nm) is from two {external-cavity diode lasers} (ECDLs) with 52~mW in total. 

\item In the cleanup region, $B^C_{0-1}$ (690~nm) is from two sets of ILAs with 90~mW of power, $A^C_{1-0}$ (637~nm) is from two sets of ILAs {with}
88~mW, and $A^C_{1-2}$ (758~nm) or $A^C_{0-1}$ (762~nm) is from a SolsTiS continuous-wave Ti:sapphire laser, with {93~mW and} a 1~GHz EOM to address the spin-rotation states.

\item In the detection region, $B^D_{0-0}$ (635~nm) is from two ECDLs with 60~mW of power, $A^D_{0-0}$ (695~nm) or $A^D_{1-1}$ (693~nm) is from two ILAs with 96~mW of power. 
\end{itemize}

The frequencies of all the transitions that we used {are}
in Table \ref{tab:transition_frequency_list}. All frequencies are measured transverse{ly} to the molecular beam, with $\le$ 10~MHz statistical uncertainties and $\le$ 60~MHz systematic uncertainties from the wavemeter. The HFS in the ground states is clearly resolved in all spectra, while the HFS in the excited states is not resolved.
Our measurements are consistent with
previous work \cite{Shayesteh_2013_CaH_fourier_spectra}.

\begin{table*}[htpb]
\centering
\begin{tabular}{|c|c|c|c|c|c|c|c|c|c|}
\hline
Ground                & $v''$                & $N''$                & $J''$                  & $F''$ & Excited                & $v'$                 & $N'$                 & $J'$                   & Frequency (THz) \\ \hline
\multirow{4}{*}{X} & \multirow{4}{*}{0} & \multirow{4}{*}{1} & \multirow{2}{*}{3/2} & 2   & \multirow{4}{*}{A} & \multirow{4}{*}{0} & \multirow{4}{*}{-} & \multirow{4}{*}{1/2} & 431.274552      \\ \cline{5-5} \cline{10-10} 
                   &                    &                    &                      & 1   &                    &                    &                    &                      & 431.274653      \\ \cline{4-5} \cline{10-10} 
                   &                    &                    & \multirow{2}{*}{1/2} & 1   &                    &                    &                    &                      & 431.276565      \\ \cline{5-5} \cline{10-10} 
                   &                    &                    &                      & 0   &                    &                    &                    &                      & 431.276512      \\ \hline
\multirow{4}{*}{X} & \multirow{4}{*}{0} & \multirow{4}{*}{1} & \multirow{2}{*}{3/2} & 2   & \multirow{4}{*}{B} & \multirow{4}{*}{0} & \multirow{4}{*}{0} & \multirow{4}{*}{1/2} & 472.026689      \\ \cline{5-5} \cline{10-10} 
                   &                    &                    &                      & 1   &                    &                    &                    &                      & 472.026790      \\ \cline{4-5} \cline{10-10} 
                   &                    &                    & \multirow{2}{*}{1/2} & 1   &                    &                    &                    &                      & 472.028702      \\ \cline{5-5} \cline{10-10} 
                   &                    &                    &                      & 0   &                    &                    &                    &                      & 472.028649      \\ \hline
\multirow{4}{*}{X} & \multirow{4}{*}{1} & \multirow{4}{*}{1} & \multirow{2}{*}{3/2} & 2   & \multirow{4}{*}{A} & \multirow{4}{*}{1} & \multirow{4}{*}{-} & \multirow{4}{*}{1/2} & 432.342011      \\ \cline{5-5} \cline{10-10} 
                   &                    &                    &                      & 1   &                    &                    &                    &                      & 432.342120      \\ \cline{4-5} \cline{10-10} 
                   &                    &                    & \multirow{2}{*}{1/2} & 1   &                    &                    &                    &                      & 432.343958      \\ \cline{5-5} \cline{10-10} 
                   &                    &                    &                      & 0   &                    &                    &                    &                      & 432.343902      \\ \hline
\multirow{4}{*}{X} & \multirow{4}{*}{1} & \multirow{4}{*}{1} & \multirow{2}{*}{3/2} & 2   & \multirow{4}{*}{B} & \multirow{4}{*}{1} & \multirow{4}{*}{0} & \multirow{4}{*}{1/2} & 471.557078      \\ \cline{5-5} \cline{10-10} 
                   &                    &                    &                      & 1   &                    &                    &                    &                      & 471.557178      \\ \cline{4-5} \cline{10-10} 
                   &                    &                    & \multirow{2}{*}{1/2} & 1   &                    &                    &                    &                      & 471.559025      \\ \cline{5-5} \cline{10-10} 
                   &                    &                    &                      & 0   &                    &                    &                    &                      & 471.558969      \\ \hline
\multirow{4}{*}{X} & \multirow{4}{*}{0} & \multirow{4}{*}{1} & \multirow{2}{*}{3/2} & 2   & \multirow{4}{*}{A} & \multirow{4}{*}{1} & \multirow{4}{*}{-} & \multirow{4}{*}{1/2} & 470.113870      \\ \cline{5-5} \cline{10-10} 
                   &                    &                    &                      & 1   &                    &                    &                    &                      & 470.113971      \\ \cline{4-5} \cline{10-10} 
                   &                    &                    & \multirow{2}{*}{1/2} & 1   &                    &                    &                    &                      & 470.115873      \\ \cline{5-5} \cline{10-10} 
                   &                    &                    &                      & 0   &                    &                    &                    &                      & 470.115819      \\ \hline
\multirow{4}{*}{X} & \multirow{4}{*}{2} & \multirow{4}{*}{1} & \multirow{2}{*}{3/2} & 2   & \multirow{4}{*}{A} & \multirow{4}{*}{1} & \multirow{4}{*}{-} & \multirow{4}{*}{1/2} & 395.717108      \\ \cline{5-5} \cline{10-10} 
                   &                    &                    &                      & 1   &                    &                    &                    &                      & 395.717218      \\ \cline{4-5} \cline{10-10} 
                   &                    &                    & \multirow{2}{*}{1/2} & 1   &                    &                    &                    &                      & 395.718978      \\ \cline{5-5} \cline{10-10} 
                   &                    &                    &                      & 0   &                    &                    &                    &                      & 395.718928      \\ \hline
\multirow{4}{*}{X} & \multirow{4}{*}{1} & \multirow{4}{*}{1} & \multirow{2}{*}{3/2} & 2   & \multirow{4}{*}{B} & \multirow{4}{*}{0} & \multirow{4}{*}{0} & \multirow{4}{*}{1/2} & 434.254840      \\ \cline{5-5} \cline{10-10} 
                   &                    &                    &                      & 1   &                    &                    &                    &                      & 434.254949      \\ \cline{4-5} \cline{10-10} 
                   &                    &                    & \multirow{2}{*}{1/2} & 1   &                    &                    &                    &                      & 434.256787      \\ \cline{5-5} \cline{10-10} 
                   &                    &                    &                      & 0   &                    &                    &                    &                      & 434.256731      \\ \hline
\multirow{4}{*}{X} & \multirow{4}{*}{1} & \multirow{4}{*}{1} & \multirow{2}{*}{3/2} & 2   & \multirow{4}{*}{A} & \multirow{4}{*}{0} & \multirow{4}{*}{-} & \multirow{4}{*}{1/2} & 393.502723      \\ \cline{5-5} \cline{10-10} 
                   &                    &                    &                      & 1   &                    &                    &                    &                      & 393.502832      \\ \cline{4-5} \cline{10-10} 
                   &                    &                    & \multirow{2}{*}{1/2} & 1   &                    &                    &                    &                      & 393.504670      \\ \cline{5-5} \cline{10-10} 
                   &                    &                    &                      & 0   &                    &                    &                    &                      & 393.504614      \\ \hline
\end{tabular}
\caption{
{F}requencies of all
transitions
used in the experiment. The \A and \B states
have unresolved hyperfine splittings. {The uncertainties are}
10~MHz statistical
and 60~MHz systematic
due to the wavemeter.
}
\label{tab:transition_frequency_list}
\end{table*}

\section{Details of measurement stages}
\label{sec:append_stages}

{The}
general
principle {for}
designing {measurement}
stages is to have at least as many independent equations as
parameters, which includes the $B$ state predissociation probability. If the number of equations and parameters are equal, as in the cases of $B(v'=0)$ and $B(v'=1)$ using method I, we can directly express \Ppd using the measured ratios. Other parameters
will also be determined and analyzed, {to}
serve as consistency checks. When there are more equations than parameters, we define a cost function to minimize the differences between the left- and right-hand sides of all equations
{(Appendix \ref{sec:bootstrapping})}.  Here we present a detailed {explanation}
of how the stages are used for predissociation {probability} measurements. We first discuss the simplest \B$(v'=0)$
measurement, where the stages include the following:

\begin{itemize}
    \item \textit{Unperturbed}.
    {Only the} $X(v''=0)\rightarrow B(v'=0)$ detection lasers are turned on. This stage serves as molecule number calibration. By taking ratios of other stages {to}
    this stage, we can eliminate molecule number $N$ from the expressions.
    \item \textit{Cleanup}. $B^C_{0-1}$ vibrational repumpers are turned on. This stage {helps to}
    estimate the $X(v''=1)$ natural population.
    \item \textit{X-A Cycling}. $A^I_{0-0}$ cycling lasers are turned on. This stage can be used to estimate the vibrational population distribution after $X-A$ cycling, and measure the depletion efficiency.
    \item \textit{X-A Cycling + Cleanup}. $A^I_{0-0}$ cycling lasers and $B^C_{0-1}$ repumps are turned on. This stage {helps to measure}
    the repump efficiency, {given}
    the $X(v''=1)$
    natural population.
    \item \textit{X-B Cycling}. $B^I_{0-0}$ cycling lasers are turned on. This stage {helps to measure the}
    vibrational population distribution after $X-B$ cycling.
    \item \textit{X-B Cycling + Cleanup}. $B^I_{0-0}$ cycling lasers and $B^C_{0-1}$ repumpers are turned on. Combined with previous stages, this {helps to measure}
    the \B state predissociation probability.
\end{itemize}

To understand the
stages better, let us take an example when $N$ ground-state molecules interact with the $X(v''=0)\rightarrow A(v'=0)$ laser. {After optical}
cycling, the downstream ground-state population
decrease{s} to $d_A N$ (where $d_A<1$ and is measurable simply by taking the ratios of signals). In this process, we
describe the depletion efficiency as $1-d_A$. We can also describe how $d_A$ is distributed among the different vibrational levels of \X
using the known VBRs. For example, the population in $X(v''=1)$
is $N (1-d_A)F_{A01}/\mathcal{F}_{A0} + n_1 N$, where $F_{A01}$, $\mathcal{F}_{A0}$ and $n_1$ represent the VBR {for}
$A(v'=0)\rightarrow X(v''=1)$, the sum of VBRs {for}
$A(v'=0)\rightarrow X(v''=1, 2, 3...)$, and normalized $X(v''=1)$ natural population, because when {a}
molecule {is}
excited to
$B(v'=0)$
it
eventually decays to a vibrational level or breaks apart. This process follows a discrete probability distribution based on the VBRs and {\Ppd.}
In the case discussed above, $(1-d_A) N$ molecules leave $B(v'=0)$,
and, based on the law of large numbers, we
expect the {$X(v''=1)$} population to become
$N (1-d_A)F_{A01}/\mathcal{F}_{A0}$.

Note that {our description relies on population transfer $(1-d_A)$ rather than the number of scattered photons.  In addition,}
the measurement protocol
does not rely on the lasers having good overlap with {the} molecular beam or with each other, because as long as molecules {share}
the same spatial and velocity distributions shot {to}
shot, the parameters (e.g., cleanup efficiency) {remain}
constant.

Here we briefly introduce the
stages in method I of {the} \B$(v'=1)$ {\Ppd}
measurement:

\begin{itemize}
    \item \textit{Unperturbed}. We always monitor the $X(v''=0)$ population,
    {which} serves as
    calibration.
    \item \textit{Cleanup}. With an $A^C_{0-1}$ laser, we
    pump the natural population in
    $X(v''=1)$ to $X(v''=0)$
    to check cleanup efficiency.
    \item \textit{State Prep}. With an $A^S_{1-0}$ 
    laser, we
    pump the natural population in $X(v''=0)$ to $X(v''=1)$
    to check state preparation efficiency. Only {after efficiently}
    pumping molecules to $X(v''=1)$
    can we perform high-SNR {optical}
    cycling on $X(v''=1)\rightarrow B(v'=1)$.
    Otherwise, the
    predissociation {loss is}
    too low to measure.
    \item \textit{State Prep + Cleanup}. We first populate $X(v''=1)$
    with $A^{S}_{1-0}$ laser, then move the $X(v''=1)$ population back to $X(v''=0)$.
    {The s}ignal {size} should be comparable to the unperturbed case. This step {helps to measure}
    $\kappa$, $a$ and $n_1$, which are cleanup efficiency, state preparation efficiency and $X(v''=1)$ natural population.
    \item \textit{State Prep + X-B 1-1 Cycling}. With most molecules in {the} $X(v''=1)$ state, we can perform optical cycling via $B(v'=1)$.
    We {expect a}
    signal increment compared to \textit{State Prep} due to {optical}
    cycling and a {re}distribution of population based on VBR and {\Ppd.}
    \item \textit{State Prep + X-B 1-1 Cycling + Cleanup}. By cleaning up the population
    in $X(v''=1)$,
    we {measure the}
    molecules
    left in $X(v''=1)$
    after {optical}
    cycling. Combined with previous stages, this provides
    5 equations and
    5 variables
    including \Ppd.
\end{itemize}

Method II is {designed} as follows. We first perform state preparation to populate {the} $X(v''=1)$ state, {similar to}
method I. By {individually} repumping the population that leaks to $X(v''=0)$ and $X(v''=2)$
we {get}
a measure of unwanted loss. This also serves as a
comparison of \A
and \B states in terms of the loss distribution. {The 10}
stages {are detailed}
in Table \ref{tab:v1 table 2}. The fact that method II accounts for
losses to {both} $X(v''=0)$ and $X(v''=2)$ {has advantages and disadvantages}
On the one hand, method II provides {an} additional {confidence}
check, with more equations than variables. Our approach to solving the over-constrained equation sets {is}
in Appendix \ref{sec:bootstrapping}.  On the other hand, {the method}
relies on detection {using}
the $X(v''=1)$ state, which {leads to lower}
signals {and higher drift sensitivity than}
detecting on $X(v''=0)$.
Hence the
SNR for
method II is not significantly {higher}
than {for} method I.

Measuring the predissociation probability for \B$(v'=2)$ and higher vibrational levels would require pumping the population to \X$(v''=2)$
and higher and performing optical cycling there, with repump{ing to}
recover the population, and monitoring unexplained loss. However, due to {practical} limitations in available space and number of lasers, as well as the increased complexity of the required stages, we {did not}
pursue these measurements.

{Table \ref{tab:full stages} contains stage details for the three types of measurement described above.} 

\begin{sidewaystable*}[htbp]
\scriptsize
\vspace{90 mm}
\centering
\begin{tabular}{ccccc}
\hline
\multicolumn{1}{|c|}{\textbf{Purpose}} &
  \multicolumn{1}{c|}{\textbf{Laser Config}} &
  \multicolumn{1}{c|}{\textbf{{\X$(v''=0)$ State Pop Normalized}}} &
  \multicolumn{1}{c|}{\textbf{{\X$(v''=1)$ State Pop Normalized}}} &
  \multicolumn{1}{c|}{\textbf{{\X$(v''=2)$ State Pop Normalized}}} \\ \hline
\multicolumn{5}{c}{$v' = 0$ experiment} \\ \hline
\multicolumn{1}{|c|}{Unperturbed} &
  \multicolumn{1}{c|}{-} &
  \multicolumn{1}{c|}{1} &
  \multicolumn{1}{c|}{$n_1$} &
  \multicolumn{1}{c|}{$\approx 0$} \\ \hline
\multicolumn{1}{|c|}{Cleanup} &
  \multicolumn{1}{c|}{$B^{C}_{0-1}$} &
  \multicolumn{1}{c|}{$1 + n_1 \kappa F_{B00}/\mathcal{F}_{B_0}$} &
  \multicolumn{1}{c|}{$ n_1 (1 - \kappa)$} &
  \multicolumn{1}{c|}{$n_1 \kappa F_{B02}/\mathcal{F}_{B_0}$} \\ \hline
\multicolumn{1}{|c|}{X-A Cycling} &
  \multicolumn{1}{c|}{$A^{I}_{0-0}$} &
  \multicolumn{1}{c|}{$d_A$} &
  \multicolumn{1}{c|}{$(1 - d_A) F_{A01} / \mathcal{F}_{A_0} + n_1$} &
  \multicolumn{1}{c|}{$(1 - d_A) F_{A02} / \mathcal{F}_{A_0}$} \\ \hline
\multicolumn{1}{|c|}{X-A Cycling + Cleanup} &
  \multicolumn{1}{c|}{$A^{I}_{0-0}  +   B^{C}_{0-1}$} &
  \multicolumn{1}{c|}{$d_A + [(1-d_A) F_{A01} / \mathcal{F}_{A_0} + n_1] \kappa F_{B00} / \mathcal{F}_{B_0}$} &
  \multicolumn{1}{c|}{$[(1-d_A) F_{A01} / \mathcal{F}_{A_0} + n_1] (1 - \kappa)$} &
  \multicolumn{1}{c|}{$[(1-d_A) F_{A01} / \mathcal{F}_{A_0} + n_1] \kappa F_{B02} / \mathcal{F}_{B_0}+(1 - d_A) F_{A02} / \mathcal{F}_{A_0}$} \\ \hline
\multicolumn{1}{|c|}{X-B Cycling} &
  \multicolumn{1}{c|}{$B^{I}_{0-0}$} &
  \multicolumn{1}{c|}{$d_B$} &
  \multicolumn{1}{c|}{$(1 - d_B) F_{B01} / \mathcal{F}_{B_1} + n_1$} &
  \multicolumn{1}{c|}{$(1 - d_B) F_{B02} / \mathcal{F}_{B_1}$} \\ \hline
\multicolumn{1}{|c|}{X-B Cycling + Cleanup} &
  \multicolumn{1}{c|}{$B^{I}_{0-0}  +  B^{C}_{0-1}$} &
  \multicolumn{1}{c|}{$d_B + [(1-d_B) F_{B01} / \mathcal{F}_{B_1} + n_1 ] \kappa F_{B00} / \mathcal{F}_{B_0}$} &
  \multicolumn{1}{c|}{$[(1-d_B) F_{B01} / \mathcal{F}_{B_1} + n_1] (1 - \kappa)$} &
  \multicolumn{1}{c|}{$[(1-d_B) F_{B01} / \mathcal{F}_{B_1} + n_1] \kappa F_{B02} / \mathcal{F}_{B_0}+(1 - d_B) F_{B02} / \mathcal{F}_{B_1}$} \\ \hline
\multicolumn{5}{c}{$v' = 1$ experiment, method I} \\ \hline
\multicolumn{1}{|c|}{Unperturbed} &
  \multicolumn{1}{c|}{-} &
  \multicolumn{1}{c|}{$1$} &
  \multicolumn{1}{c|}{$n_1$} &
  \multicolumn{1}{c|}{$\approx 0$} \\ \hline
\multicolumn{1}{|c|}{State Prep} &
  \multicolumn{1}{c|}{$A^{S}_{1-0}$} &
  \multicolumn{1}{c|}{$1 - a$} &
  \multicolumn{1}{c|}{$n_1 + a F_{A11}/\mathcal{F}_{A_2}$} &
  \multicolumn{1}{c|}{$a F_{A12}/\mathcal{F}_{A_2}$} \\ \hline
\multicolumn{1}{|c|}{Cleanup} &
  \multicolumn{1}{c|}{$A^{C}_{0-1}$} &
  \multicolumn{1}{c|}{$1 + n_1 \kappa F_{A00}/\mathcal{F}_{A_1}$} &
  \multicolumn{1}{c|}{$n_1 (1 - \kappa)$} &
  \multicolumn{1}{c|}{$n_1 \kappa F_{A02}/\mathcal{F}_{A_1}$} \\ \hline
\multicolumn{1}{|c|}{State Prep + Cleanup} &
  \multicolumn{1}{c|}{$A^{S}_{1-0}  +  A^{C}_{0-1}$} &
  \multicolumn{1}{c|}{$1 - a + (n_1 + a F_{A11} / \mathcal{F}_{A_2}) \kappa F_{A00} / \mathcal{F}_{A_1}$} &
  \multicolumn{1}{c|}{$(n_1 + a F_{A11}/\mathcal{F}_{A_2}) (1 - \kappa)$} &
  \multicolumn{1}{c|}{$a F_{A12}/\mathcal{F}_{A_2} + (n_1 + a F_{A11} / \mathcal{F}_{A_2}) \kappa F_{A02} / \mathcal{F}_{A_1}$} \\ \hline
\multicolumn{1}{|c|}{State Prep + X-B 1-1 Cycling} &
  \multicolumn{1}{c|}{$A^{S}_{1-0}  +   B^{I}_{1-1}$} &
  \multicolumn{1}{c|}{$1 - a + (n_1 + a F_{A11} / \mathcal{F}_{A_2}) d_B F_{B10} / \mathcal{F}_{B_2}$} &
  \multicolumn{1}{c|}{$(n_1 + a F_{A11}/\mathcal{F}_{A_2}) (1 - d_B)$} &
  \multicolumn{1}{c|}{$a F_{A12}/\mathcal{F}_{A_2} + (n_1 + a F_{A11} / \mathcal{F}_{A_2}) d_B F_{A02} / \mathcal{F}_{A_1}$} \\ \hline
\multicolumn{1}{|c|}{\begin{tabular}[c]{@{}c@{}}State Prep + X-B 1-1 Cycling\\  + Cleanup\end{tabular}} &
  \multicolumn{1}{c|}{$A^{S}_{1-0}  +   B^{I}_{1-1}  +  A^{C}_{0-1}$} &
  \multicolumn{1}{c|}{\begin{tabular}[c]{@{}c@{}}$1 - a + (n_1 + a F_{A11} / \mathcal{F}_{A_2}) [d_B F_{B10} / \mathcal{F}_{B_2}$\\ $+ (1 - d_B) \kappa F_{A00} / \mathcal{F}_{A_1}]$\end{tabular}} &
  \multicolumn{1}{c|}{$(n_1 + a F_{A11}/\mathcal{F}_{A_2}) (1 - \kappa) (1 - d_B)$} &
  \multicolumn{1}{c|}{\begin{tabular}[c]{@{}c@{}}$a F_{A12}/\mathcal{F}_{A_2} + (n_1 + a F_{A11} / \mathcal{F}_{A_2}) [d_B F_{B12} / \mathcal{F}_{B_2}$\\ $+ (1 - d_B) \kappa F_{A02} / \mathcal{F}_{A_1}]$\end{tabular}} \\ \hline
\multicolumn{5}{c}{$v' = 1$ experiment, method II} \\ \hline
\multicolumn{1}{|c|}{State Prep + Cleanup v0} &
  \multicolumn{1}{c|}{$A^{S}_{1-0} + A^{C}_{1-0}$} &
  \multicolumn{1}{c|}{$(1 - a) (1 - \kappa_1)$} &
  \multicolumn{1}{c|}{$n_1 + (a + \kappa_1 - a \kappa_1) F_{A11}/\mathcal{F}_{A_2}$} &
  \multicolumn{1}{c|}{$(a + \kappa_1 - a \kappa_1) F_{A12}/\mathcal{F}_{A_2}$} \\ \hline
\multicolumn{1}{|c|}{Unperturbed} &
  \multicolumn{1}{c|}{-} &
  \multicolumn{1}{c|}{$1$} &
  \multicolumn{1}{c|}{$n_1$} &
  \multicolumn{1}{c|}{$\approx 0$} \\ \hline
\multicolumn{1}{|c|}{State Prep} &
  \multicolumn{1}{c|}{$A^{S}_{1-0}$} &
  \multicolumn{1}{c|}{$1 - a$} &
  \multicolumn{1}{c|}{$n_1 + a F_{A11}/ \mathcal{F}_{A_2} \equiv \mathcal{Z}$} &
  \multicolumn{1}{c|}{$a F_{A12}/\mathcal{F}_{A_2}$} \\ \hline
\multicolumn{1}{|c|}{Cleanup v0} &
  \multicolumn{1}{c|}{$A^{C}_{1-0}$} &
  \multicolumn{1}{c|}{$1 - \kappa_1$} &
  \multicolumn{1}{c|}{$n_1 + \kappa_1 F_{A11}/\mathcal{F}_{A_2}$} &
  \multicolumn{1}{c|}{$\kappa_1 F_{A12}/\mathcal{F}_{A_2}$} \\ \hline
\multicolumn{1}{|c|}{State Prep + X-A 1-1 Cycling} &
  \multicolumn{1}{c|}{$A^{S}_{1-0} + A^{I}_{1-1}$} &
  \multicolumn{1}{c|}{$1 - a + \mathcal{Z} d_A F_{A10}/ \mathcal{F}_{A_3}$} &
  \multicolumn{1}{c|}{$\mathcal{Z} (1 - d_A)$} &
  \multicolumn{1}{c|}{$a F_{A12}/\mathcal{F}_{A_2} + \mathcal{Z} d_A F_{A12} / \mathcal{F}_{A_3}$} \\ \hline
\multicolumn{1}{|c|}{\begin{tabular}[c]{@{}c@{}}State Prep + X-A 1-1 Cycling\\  + Cleanup v0\end{tabular}} &
  \multicolumn{1}{c|}{$A^{S}_{1-0} +  A^{I}_{1-1} + A^{C}_{1-0}$} &
  \multicolumn{1}{c|}{$(1 - a + \mathcal{Z} d_A F_{A10}/ \mathcal{F}_{A_3}) (1 - \kappa_1)$} &
  \multicolumn{1}{c|}{\begin{tabular}[c]{@{}c@{}}$\mathcal{Z} (1 - d_A) + (1 - a + \mathcal{Z} d_A F_{A10} / \mathcal{F}_{A_3}) $\\ $\kappa_1 F_{A11} / \mathcal{F}_{A_2}$\end{tabular}} &
  \multicolumn{1}{c|}{\begin{tabular}[c]{@{}c@{}}$a F_{A12}/\mathcal{F}_{A_2} + \mathcal{Z} d_A F_{A12} / \mathcal{F}_{A_3}$\\ $ + (1 - a + \mathcal{Z} d_A F_{A10} / \mathcal{F}_{A_3}) \kappa_1 F_{A12} / \mathcal{F}_{A_2}$\end{tabular}} \\ \hline
\multicolumn{1}{|c|}{\begin{tabular}[c]{@{}c@{}}State Prep + X-A 1-1 Cycling\\  + Cleanup v2\end{tabular}} &
  \multicolumn{1}{c|}{$A^{S}_{1-0} +  A^{I}_{1-1} + A^{C}_{1-2}$} &
  \multicolumn{1}{c|}{\begin{tabular}[c]{@{}c@{}}$1 - a + \mathcal{Z} d_A F_{A10}/ \mathcal{F}_{A_3} + (a F_{A12}/\mathcal{F}_{A_2} $\\ $+ \mathcal{Z}d_A F_{A12}/\mathcal{F}_{A_3})\kappa_2 F_{A10}/\mathcal{F}_{A_4}$\end{tabular}} &
  \multicolumn{1}{c|}{\begin{tabular}[c]{@{}c@{}}$\mathcal{Z} (1 - d_A) + (a F_{A12} / \mathcal{F}_{A_2} $\\ $+ \mathcal{Z} d_A F_{A12} / \mathcal{F}_{A_3}) \kappa_2 F_{A11} / \mathcal{F}_{A_4}$\end{tabular}} &
  \multicolumn{1}{c|}{$(a F_{A12}/\mathcal{F}_{A_2} + \mathcal{Z} d_A F_{A12} / \mathcal{F}_{A_3}) (1 - \kappa_2)$} \\ \hline
\multicolumn{1}{|c|}{State Prep + X-B 1-1 Cycling} &
  \multicolumn{1}{c|}{$A^{S}_{1-0} + B^{I}_{1-1}$} &
  \multicolumn{1}{c|}{$1 - a + \mathcal{Z} d_B F_{B10}/ \mathcal{F}_{B_2}$} &
  \multicolumn{1}{c|}{$\mathcal{Z} (1 - d_B)$} &
  \multicolumn{1}{c|}{$a F_{A12}/\mathcal{F}_{A_2} + \mathcal{Z} d_A F_{B12} / \mathcal{F}_{B_2}$} \\ \hline
\multicolumn{1}{|c|}{\begin{tabular}[c]{@{}c@{}}State Prep + X-B 1-1 Cycling\\  + Cleanup v0\end{tabular}} &
  \multicolumn{1}{c|}{$A^{S}_{1-0} +  B^{I}_{1-1} + A^{C}_{1-0}$} &
  \multicolumn{1}{c|}{$(1 - a + \mathcal{Z} d_B F_{B10}/ \mathcal{F}_{B_2}) (1 - \kappa_1)$} &
  \multicolumn{1}{c|}{\begin{tabular}[c]{@{}c@{}}$\mathcal{Z} (1 - d_B) + (1 - a + \mathcal{Z} d_B F_{B10} / \mathcal{F}_{B_2}) $\\ $\kappa_1 F_{A11} / \mathcal{F}_{A_2}$\end{tabular}} &
  \multicolumn{1}{c|}{\begin{tabular}[c]{@{}c@{}}$a F_{A12}/\mathcal{F}_{A_2} + \mathcal{Z} d_B F_{B12} / \mathcal{F}_{B_2}$\\ $ + (1 - a + \mathcal{Z} d_B F_{B10} / \mathcal{F}_{B_2}) \kappa_1 F_{A12} / \mathcal{F}_{A_2}$\end{tabular}} \\ \hline
\multicolumn{1}{|c|}{\begin{tabular}[c]{@{}c@{}}State Prep + X-B 1-1 Cycling\\  + Cleanup v2\end{tabular}} &
  \multicolumn{1}{c|}{$A^{S}_{1-0} +  B^{I}_{1-1} + A^{C}_{1-2}$} &
  \multicolumn{1}{c|}{\begin{tabular}[c]{@{}c@{}}$1 - a + \mathcal{Z} d_B F_{B10}/ \mathcal{F}_{B_2} + (a F_{A12}/\mathcal{F}_{A_2} $\\ $+ \mathcal{Z}d_B F_{B12}/\mathcal{F}_{B_2})\kappa_2 F_{A10}/\mathcal{F}_{A_4}$\end{tabular}} &
  \multicolumn{1}{c|}{\begin{tabular}[c]{@{}c@{}}$\mathcal{Z} (1 - d_B) + (a F_{A12} / \mathcal{F}_{A_2} $\\ $+  \mathcal{Z} d_B F_{B12} / \mathcal{F}_{B_2}) \kappa_2 F_{A11} / \mathcal{F}_{A_4}$\end{tabular}} &
  \multicolumn{1}{c|}{$(a F_{A12}/\mathcal{F}_{A_2} + \mathcal{Z} d_B F_{B12} / \mathcal{F}_{B_2}) (1 - \kappa_2)$} \\ \hline
\end{tabular}
\caption{Complete stage information for {the} \B state predissociation measurement {(also summarized in Tables \ref{tab:v0 table}, \ref{tab:v1 table 1}, and \ref{tab:v1 table 2})}.
For $B(v'=0)$ we use 5 variables:
$n_1$, $\kappa$, $F_{B0a}$, $d_A$, and $d_B$, representing $X(v''=1)$
natural population, cleanup efficiency of laser $B^{C}_{1-0}$, {the} $B(v'=0)$ state predissociation probability, depletion efficiency of laser $A^{I}_{0-0}$, and depletion efficiency of laser $B^{I}_{0-0}$.
For $B(v'=1)$ method I, {the 5}
variables
used {are}
$a$, $n_1$, $\kappa$, $F_{B1a}$, and $d_B$, representing state preparation (from $X(v''=0)$ to $X(v''=1)$) efficiency, $X(v''=1)$
natural population, cleanup efficiency of laser $A^{C}_{1-0}$, {the} $B(v'=1)$ state predissociation probability, {and} depletion efficiency of laser $B^{I}_{1-1}$.
For $B(v'=1)$ method II{, the 7}
variables {are}
$a$, $n_1$, $\kappa_1$, $\kappa_2$, $F_{B1a}$, $d_A$, and $d_B$, representing state preparation (from $X(v''=0)$ to $X(v''=1)$) efficiency, $X(v''=1)$
natural population, cleanup efficiency of laser $A^{C}_{1-0}$, cleanup efficiency of laser $A^{C}_{1-2}$, {the} $B(v'=1)$ state predissociation probability, depletion efficiency of laser $A^{I}_{1-1}$, {and} depletion efficiency of laser $B^{I}_{1-1}$.
{We denote the VBR normalization factors as $\mathcal{F}_{A_0} \equiv \sum_{i\neq0}F_{A0i}$, $\mathcal{F}_{A_1} \equiv \sum_{i\neq1}F_{A0i}$, $\mathcal{F}_{A_2} \equiv \sum_{i\neq0}F_{A1i}$, $\mathcal{F}_{A_3} \equiv \sum_{i\neq1}F_{A1i}$, $\mathcal{F}_{A_4} \equiv \sum_{i\neq2}F_{A1i}$, $\mathcal{F}_{B_0} \equiv F_{B0a} + \sum_{i\neq1}F_{B0i}$,  $\mathcal{F}_{B_1} \equiv F_{B0a} + \sum_{i\neq0}F_{B0i}$, and $\mathcal{F}_{B_2} \equiv F_{B1a} + \sum_{i\neq1}F_{B1i}$.}
}
\label{tab:full stages}
\end{sidewaystable*}

\section{Bootstrap method used in the predissociation data analysis}
\label{sec:bootstrapping}

Bootstrapping is a statistical technique that involves generating multiple samples from a dataset by sampling with replacement \cite{Efron_1979_bootstrapping}. It is a useful tool for constructing confidence intervals for a population parameter, {$-$}
in {this}
case,
the expectation values of predissociation probabilities. 

{A key}
benefit of bootstrapping is that it allows one to make inferences about a population based on a sample, without making any assumptions about the underlying distribution of the population. Given the complexity of the functional form of predissociation probability with respect to experimentally measured ratios, utilizing {a} bootstrap
method {helps to}
avoid {assuming a}
normal distribution when determining the confidence interval of predissociation probabilities.

One way to {use such a method on}
a set of data $d$ with size $n$ is to use the array of $d$ data points to generate $n$ ``bootstrapped'' samples by sampling with replacement. We can then compute a statistic of interest, such as the mean, from {the} $n$ bootstrapped samples, and save {it}
to a new array. We repeatedly generate $n$ bootstrapped samples, calculate the mean, and save it to the storage array. The resulting distribution of the statistic can then be used to make inferences about the population.

Let us {consider}
the $B(v'=0)$ predissociation probability as an example. The experimental procedure to acquire ratios is shown in Fig.~\ref{fig:graphical_ratios} and explained in Sec.~\ref{subsec:result_analysis}. All the ratios ($r_1,\;r_2,\;r_3,\;r_4,\;r_5$) are expressed using the variables
in Table \ref{tab:v0 table}, including $n_1$, $\kappa$, $F_{B0a}$, $d_A$ and $d_B$. Five equations can be explicitly written as
\begin{equation} \label{eq:ri}
\begin{split}
r_1 &= 1 + n_1 \kappa F_{B00}/\mathcal{F}_{B_0}, \\
  r_2 &= d_A, \\
  r_3 &= d_A + [(1-d_A) F_{A01} / \mathcal{F}_{A_0} + n_1] \kappa F_{B00} / \mathcal{F}_{B_0}, \\
  r_4 &= d_B, \\
  r_5 &= d_B + [(1-d_B) F_{B01} / \mathcal{F}_{B_1} + n_1 ] \kappa F_{B00} / \mathcal{F}_{B_0}.
\end{split}
\end{equation}

By solving these 5 equations {for 5}
variables, we {can}
express $F_{B0a}$ as a function of $r_i$ ($i \in \{1,2,3,4,5\}$) with known VBRs. Therefore, we {obtain}
a function $f_{B0a}$ that takes in $r_i$ ($i \in \{1,2,3,4,5\}$) and outputs predissociation probability $F_{B0a}$. Here we describe the procedure of performing bootstrap analysis on the data, where the data {consists of}
$n\approx200$ sets of ratios $\{ r_1, r_2, r_3, r_4, r_5 \}$, with $r_i$ ($i \in \{1,2,3,4,5\}$) being an array of length $n$:

\begin{enumerate}
    \item Randomly sample $n$ elements from the original $r_1$ array with replacement, {i.e.,}
    elements from {the} original $r_1$
    can appear more than once in the new array $r_1^{\mathrm{bt}}$. This step mimics the situation where the same measurement is performed again. We {carry out}
    independent random sampling with replacement for $r_2$, $r_3$, $r_4$, and $r_5$ as well, and {obtain}
    $r_i^{\mathrm{bt}}$ ($i \in \{1,2,3,4,5\}$) arrays.
    \item Calculate the mean of the newly generated $r_i^{\mathrm{bt}}$ arrays individually, which can be denoted as $\overline{r_i^{\mathrm{bt}}}$. We can feed these $\overline{r_i^{\mathrm{bt}}}$s to the function $f_{B0a}$ and store the output in an array $F$.
    \item Repeat steps 1 and 2 for {$\sim10^6$}
    times, until the statistical properties such as mean and standard deviation of the normalized distribution of array $F$ converge.
    \item Analyze the distribution of $F$. For the expectation value, we use the mean of array $F$. To determine the 95\% confidence interval, we take the 2.5\% quantile from the distribution of $F$ as the lower bound, and {the} 97.5\% quantile as the upper bound.
\end{enumerate}

The data analysis for $B(v'=1)$ method I is almost identical to that for $B(v'=0)$. The {bootstrap} procedures
are the same, and
the analysis code can be found online\footnote{github.com/QiSun97/CaH\_Predissociation/bootstrapping\_v6\_final\_github.ipynb}. 

The data analysis for $B(v'=1)$ method II is slightly different from the previous two cases. We no longer have a deterministic function of $F_{B1a}$ because {there are 9}
equations {with 7}
variables.
To solve this over-constrained system,
we
perform a least square fit. We
write down the 9 equations {with}
all the terms on the right hand side {and zeros on the}
left hand side. {Then}
we define the cost function as the sum of squares of all the right hand sides of the equations, and use {the} Levenberg-Marquardt algorithm to search for the local minimum with a reasonable initial guess.

\section{Theoretical details}
\label{sec:nacme}

The following three-state Hamiltonian (for the \X, \B and $D^2\Sigma^+$\xspace electronic states) is diagonalized to obtain wavefunctions, FCFs, and predissociation estimates:
\begin{equation}
\sum^{3}_{j}H_{j} = \sum_{I}\frac{(\hat{p}^{I}+A^{I}_{ij}(r))^{2}}{2\mu}\phi_{j}(r)+V_{j}(r)\phi_{j}(r).
\end{equation}
The first term is the kinetic energy operator, in which $\hat{p}$ is the standard momentum operator, expressed on a grid via the Colbert-Miller derivative.  We represent the momentum operator in position space so that we can incorporate the nonadiabatic coupling term directly. {}This term
is computed in the position representation, $A_{ij}(r) = \braket{\Psi_{i}|\hat{p}\Psi_{j}}$.
We obtain $\braket{\Psi{i}|\frac{\partial}{\partial r}\Psi_{j}}$ from a $dr = 0.001a_0$ potential energy surface scan via Molpro electronic structure calculations, and interpolate this onto a spline to represent $A_{ij}(r)$.
The reduced mass of CaH {is $\mu$}.  Finally, $V(r)$ is obtained from the $dr = 0.001a_0$ scan via the MRCI+Q Davidson energies and interpolated onto a spline before being incorporated into the Hamiltonian.  
 
{At $r_0=8$
a.u., an optical potential of the form $-iV(r-r_0)^{2}/w^{2}$ is added only to the \X state's $V(r)$ at the PES asymptote with each grid-point to simulate the continuum and create a flux equation. 
Specifically, the optical absorbing potential must have a width $w$ and depth $V$ which guarantees complete wavepacket absorption and ensures the potential is smooth so that hardly any reflection takes place before the wavepacket enters the potential \cite{Neuhauser_1989_absorber}.  The absorber width is chosen to be $w = 8~a_0$, much larger than the typical de Broglie wavelength of $\sim 0.2~a_0$. We choose a depth as the typical energy of the wavepacket, or 0.2 a.u. ($4.4\times10^4$~cm$^{-1}$).}
The Hamiltonian is then diagonalized.  The optical potential enforces imaginary eigenvalues {that}
are directly related to
nonradiative {loss} rates, which are then compared to the radiative rates calculated from the MRCI-computed transition dipole moments to obtain a predissociation probability.

For spin-orbit coupling, the active space for \X, \A and \B states must be the same, therefore a compromise is chosen to optimize the \X and \B FCFs over the \A.
Interestingly, we note that using our basis set and active space but shifting the \A potential energy surface can
produce FCFs that are equivalent to experimentally measured {values}, as {shown}
in Table \ref{tab:PESshift}. This is because static electron correlation has converged, but important dynamic correlation is missing. This depends on the original orbital active space from CASSCF which then affects the MRCI equilibrium bond length. 

\begin{table}[htpb]
\centering
\begin{tabular}{|c|c|c|c|}
\hline
Transition & \begin{tabular}[c]{@{}c@{}}Vibrational Quanta\\ $(v'')$\end{tabular} & \begin{tabular}[c]{@{}c@{}}FCF Calculated \\ $(f_{0v''})$\end{tabular} & \begin{tabular}[c]{@{}c@{}}FCF Measured\\ $(f_{0v''})$\end{tabular} \\ \hline
\multirow{4}{*}{$A'\rightarrow X$} & 0 & 0.9568 & 0.9572(43) \\ \cline{2-4} 
 & 1 & 0.0401 & 0.0386(32) \\ \cline{2-4} 
 & 2 & 2.9$\times 10^{-3}$ & 4.2(3.2)$\times 10^{-3}$ \\ \cline{2-4} 
 & 3 & 2.5$\times 10^{-4}$ & - \\ \hline
\end{tabular}
\caption{Calculated and measured values of FCFs for CaH. We show experimental FCFs
\cite{Vasquez-Carson_2022_CaH} for comparison. $A'$ is the same active space as the \B state in this work, shifting the potential to the left by $0.0375a_0$.}
\label{tab:PESshift}
\end{table}

\end{document}